# Recent Advances in Two-Dimensional Metal Monochalcogenides

Abdus Salam Sarkar[1]* and Emmanuel Stratakis[1,2]*

[1]Institute of Electronic Structure and Laser, Foundation for Research and Technology-Hellas, Heraklion, 700 13 Crete, Greece.

[2]Physics Department, University of Crete, Heraklion, 710 03 Crete, Greece.

**Email:** salam@iesl.forth.gr; stratak@iesl.forth.gr

**Graphical ToC**

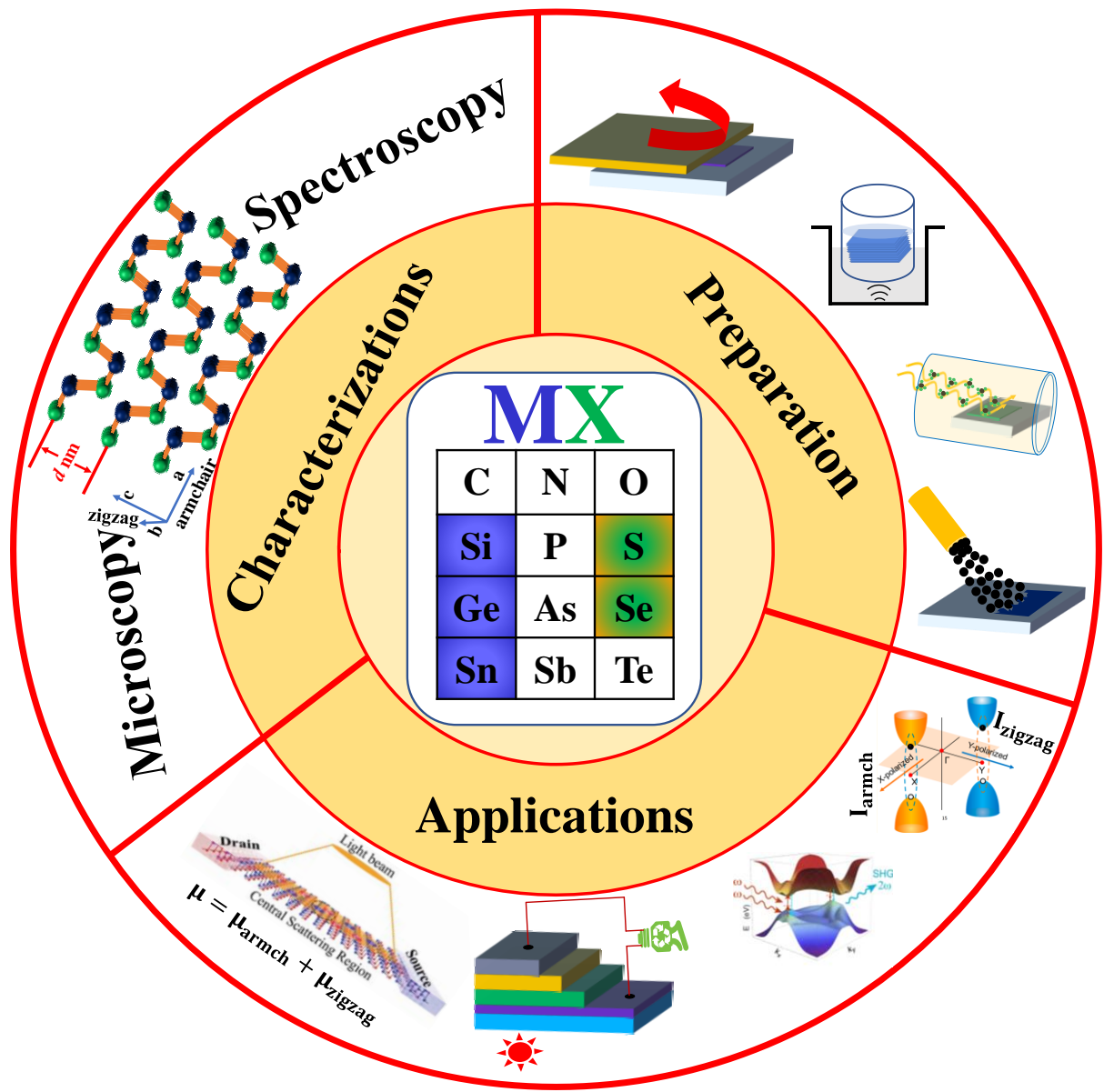

**Abstract**

The family of emerging low-symmetry and structural in-plane anisotropic 2D materials have been expanding rapidly in recent years. As an important emerging anisotropic 2D material, the black phosphorene (BP) analog group $IV_A$-VI metal monochalcogenides (MMCs) have been surged recently due to their distinctive crystalline symmetries, exotic in-plane anisotropic electronic and optical response, earth abundant and environmentally friendly characteristics. In this article, we review the recent research advancements in the field of anisotropic 2D MMCs. At first, the unique wavy crystal structures together with the optical and electronic properties of such materials are discussed. The review continues with the various methods adopted for the synthesis of layered MMCs including micromechanical and liquid phase exfoliation as well as physical vapor deposition. The last part of the article focuses on the application of the structural anisotropic response of 2D MMCs in field effect transistors, photovoltaic cells non-linear optics and valleytronic devices. Besides presenting the significant research in the field of this emerging class of 2D materials, this review also delineates the existing limitations and discusses emerging possibilities and future prospects.

# 1. Introduction

Since the successful isolation of graphene,[1] research in atomically thin two dimensional (2D) materials has gained intensive interest. In particular, beyond graphene materials such as hexagonal boron nitride (h-BN),[2-4] transition metal dichalcogenides (TMDs),[2, 4-6] multinary layered chalcogenides,[7] perovskites,[8-10] and MXenes[11-13] have opened up a new horizon in 2D material research. These 2D materials possess intriguing optical, electronic, mechanical and optoelectronic properties, and are being explored for rich physics and many emerging scientific applications.[1, 3-5, 9, 12, 14, 15] However, completing the intensive research on semimetal graphene and beyond graphene semiconducting 2D materials, group V element black phosphorene (BP) discovered as a 2D material in 2014,[16] leading to many discoveries of novel physical phenomena.[16-20] The puckered or wavy lattice structure with reduced crystal symmetry ($D_{2h}$) than graphene ($D_{6h}$) and TMDs make it more interesting for exhibiting the novel physical phenomena. Moreover, layer dependent tunable bandgap,[21] high carrier mobility,[16, 20] and strong in-plane anisotropy[18] make it promising for next generation emerging electronic and photonic applications. The electronic band gap of BP varies from 0.33 to 2.0 eV, when the thickness reduced to a monolayer. An ultrathin layer of BP revealed an extraordinary electrical hole mobility ~$5 \times 10^3$ cm$^2$ V$^{-1}$ s$^{-1}$ at room temperature,[22] which is much higher than the TMDs,[23] making it suitable for advanced electronic applications. Most importantly, in-plane anisotropic optical and electrical response along *x*-direction (armchair) and *y*-direction (zig-zag) was directly visualized and studied, which added a new dimension.[18, 24] Beside this, unique structural in-plane anisotropic nature in BP played a critical role in designing multifunctional and tunable 2D novel electronic, optoelectronic and photonic devices.[17, 18] In spite of a series of exotic novel physical phenomena, rapid ambient degradation of phosphorene is a critical issue for its practical implementations.[25-27]

Most interestingly, beyond graphene 2D materials, BP's isostructural and isoelectronic group IV$_A$-VI metal monochalcogenides (MMCs), with chemical formula MX (M= Si, Ge and Sn and X=chalcogens) has been surged as a star 2D materials (**Figure 1**) due to the low cost, earth abundant and environmentally friendly features.[28-33] In 2015, first theoretical and experimental works revealing the electronic properties of monolayer or few-layer MMCs were published.[31, 34] The results, showed a direct and indirect band gap with 1.0 to 2.3 eV energies covering part of the infrared and visible range, which is higher than the BP. Thereafter, the phonon-limited electronic carrier mobilities of MX monolayers are estimated theoretically to be on the order of $10^3$ to $10^5$ cm$^2$ V$^{-1}$s$^{-1}$ by Xu et al.[35] Highest anisotropic electronic response ratio in a few layer MXs has been recorded along armchair and zigzag direction to be ~5.8,[36] which is larger than the existing anisotropic 2D materials. Most notably, monolayer MMCs exhibits orthorhombic crystal structure (Pnma space group) with low crystal symmetry C$_{2v}$, in which the inversion symmetry is further broken. Which makes them possible to observe new order parameters (spin orbital coupling) and polarization properties. Many exotic phenomena have been predicted on a monolayer MMCs, including valley physics,[37] spontaneous polarization and bulk photovoltaic effect,[38-41] piezo-phototronic,[42] giant piezoelectricity,[43] ferroelectricity,[44-47] multiferroics of ferroelectricity,[48] and ferroelasticity.[49] However, valley selective dichroism,[50] and more than 90% room temperature valley polarization degree[51] in SnS provide a completely novel platform for valleytronics. In addition, the giant optical second harmonic generation (SHG) in MMCs are also promising in nonlinear optoelectronic applications.[46, 52]

In this article, the latest advances in the field of the emerging 2D group IVA-VI metal monochalcogenide materials (MMCs), are reviewed. The intriguing physical (crystal and electronic structure) and optical properties are initially highlighted and discussed. Then the various methods that have been employed for the synthesis of such materials, including mechanical and liquid phase exfoliation, as well as vapor phase deposition techniques are demonstrates. Besides presenting the potential and significance of 2D MMCs in various electronic applications it will also delineate the existing limitations and discuss emerging possibilities and future prospects.

## 2. Crystal structure and properties of MMCs

### 2.1 Crystal structure of MMCs

The monolayer crystal structure of group IV$_A$-VI metal monochalcogenides (MMCs) is isostructural with orthorhombic black phosphorus.[28, 31] The chemical formula of MMCs is the MX, where, M is group IV metals (Si, Ge, Sn) and X is chalcogens (S and Se). M and X atoms are alternate with each other and form a puckered or wavy layer structure of zigzag (*y*) and the armchair (*x*) plane (**Figure 1a, b**).[28, 31, 41, 53] In MXs, the presence of two atomic species (M and X) with different electronegativity lowers the crystal symmetry compared to other 2D crystals. In particular, the bulk structure belongs to the space group Pnma($D_{2h}^{16}$) , while in the monolayer, the inversion symmetry is broken placing them in the $D_{2v}^{7}$ space group. The MXs monolayer has four atoms per unit cell (**Figure 1b, c**), in which each atom is covalently bonded to three neighbors of the other, forming zigzag (*y*) rows of alternating elements. The corresponding first Brillouin zone, as well as the high-symmetry points along armchair (*x*) and zig-zag (*y*) directions, are indicated in **Figure 1d**. The valence electronic configuration of metal (M) and chalcogens (X) atoms are $4d^{10}ns^2np^2$ and $ns^2np^4$ where, $n$ is the periodic number of the element. MMCs show strong covalent bonding in the 2D plane and a strong interlayer force owing to the lone pair electrons, which generate a large electron distribution and electronic coupling between adjacent layers.[54-57] The MMCs are layered p-type semiconductors with 1:1 stoichiometry, with most studied examples be SiS, SiSe, GeS, GeSe,

SnS, and SnSe. In particular, SnS has typical acceptor states, which are formed by Sn vacancies ($V_{Sn}$). On the other hand, sulfur vacancies ($V_s$) can be also formed, under appropriate Sn-rich conditions, as well as substitutional oxygen at sulfur sites ($O_S$), leading, in all cases, to p-type conductivity.[58] Stereochemically, active lone pair electrons in group IV metals electronic configuration, $ns$ ($n$=5 or 4), play a pivotal role in the structural distortion, resulting in the anisotropic crystal and layered structure.[57, 59, 60]

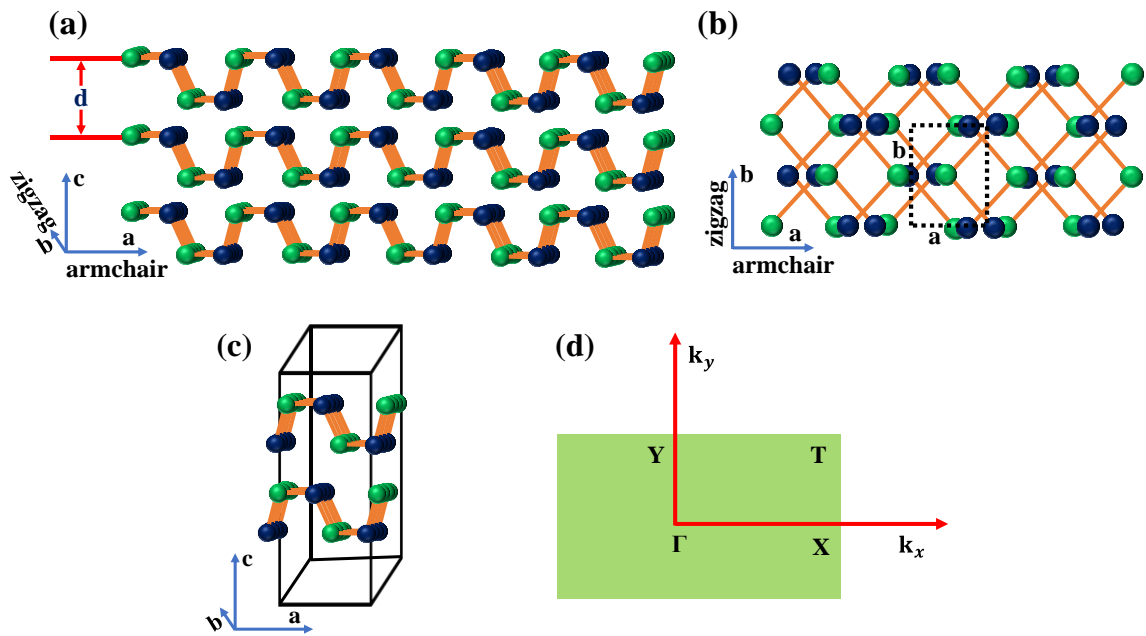

**Figure 1.** *Atomic structures of phosphorene analogous MXs a) side view and b) top view with the lattice vectors a and b along the x (armchair) and y (zig-zag) directions. Green atoms are metal (M=Si, Ge and Sn) and blue atoms are group VI chalcogens (X=S and Se). Each single layer is of thickness of d nm. c) The orthorhombic unit cell of MXs. and d) the first Brillouin zone with high-symmetry points Γ, X, T, and Y $k_x$, $k_y$ momentum along armchair and zig-zag direction.*

The physical properties of the MMCs are closely relative to their crystal structure, which immensely influences the optical and electronic properties. Indeed, MXs exhibit various crystal phases, such as orthorhombic, hexagonal and cubic. The exact phase is controlled by the different oxidation states of the metal and chalcogen atoms. For example in Sn-based MMCs, the chalcogen atom has a stronger electronegativity than the metal. As a result the chalcogen captures two atoms from Sn, which leads to a change in its electronic configuration from $4d^{10}5s^25p^2$ to $4d^{10}5s^25p^0$; the same is true for Se-based MMCs, where, the electronic configuration of Se changes to $4s^24p^6$.[54, 61] As a consequence, the buckled crystal layer structure is distorted (**Figure 1**). The typical physical properties of the MMCs reported to date are presented in **Table 1**.[62-72]

**Table 1.** *Summary of typical MMCs crystal structure parameters and their properties.*

| MMCs | Space group | Crystal structures | Lattice parameters (Å) | Distance $d$ (nm) | References |
|------|-------------|--------------------|------------------------|-------------------|------------|
| SiS | | Orthorhombic | a=4.61; b=3.27 | 0.228-0.234 | *[62, 63]* |
| SiSe | | Orthorhombic | b=5.0; b=3.54 | 0.246-0.249 | *[63, 64]* |
| GeS | $D_{2h}^{16}$ | Orthorhombic | a=4.3; b=10.47; c=3.65 | 0.56 | *[65-67]* |
| GeSe | $D_{2h}^{16}$ | Orthorhombic | a=10.84; b=3.83; c=4.39 | - | *[67-69]* |
| SnS | $D_{2h}^{16}$ | Orthorhombic | a=4.33; b=11.19; c=3.98 | 0.56 | *[70]* |
| SnSe | $D_{2h}^{16}$ | Orthorhombic | a=11.49; b=4.15; c=4.44 | 0.62 | *[71, 72]* |

*distance $d$ is thickness of a monolayer MXs.

The exotic physics of 2D materials is usually associated with crystal structural symmetry breaking. In particular, 2D graphene has highest symmetry of $D_{6h}$, which has six fold rotation (in-plane), six two-fold perpendicular axis and a mirror plane. A puckered structure of monolayer black phosphorus exhibits $D_{2h}$ symmetry, which comprises a two-fold rotation axes and one mirror plane, leading to highly anisotropic optical, electronic and thermal properties.[24,

[73-77] However, MMCs consist of two elements with different electronegativity, as opposed to the single element in black phosphorus. As a result, inversion symmetry in odd layer is broken to $C_{2v}$, which, in addition to a two-fold rotation, it contains two mirror planes.[43] This unique feature leads to even more extraordinary optical and electronic properties than that of phosphorene and TMDs.[37, 39, 43, 44, 47, 51, 78] Apart from such intrinsic symmetries, other external factors can further tune the electronic properties, including piezo-phototronic and photoactivity.

The bulk structure of Ge and Sn based monochalcogenides has a layered orthorhombic crystal structure of the space group Pnma ($D_{2h}^{16}$). The structure exhibits strong interlayer forces, giving rise to a distorted NaCl (d-NaCl) crystal structure. As a result, the crystals have different perspective views along a, b, and c axial directions, presenting a unique anisotropic nature. On the other hand, the Si based monochalcogenides belong to different space groups, for example the α-SiS monolayer structure.[79]

## 2.2 Band structure, optical and carrier transport properties

The electronic band structure of 2D materials is crucial for the understanding of the electronic and optical processes occurred in versatile electronic and photonic device applications. In particular, the layer dependent electronic band structure of phosphorene-analogue MMCs has been investigated by Gomes et al.[31] A tunable band-gap ($E_g$) with direct or indirect band gap energy within the visible range is reported (**Table 2**). The electronic structure and properties were obtained by calculating the band structures using the *Heyd-Scuseria-Ernzerhof (*HSE) function. On the other hand, ab initio density functional theory using the Perdew-Burke-Ernzerhof (PBE) exchange correlation function, is used for SiS and SiSe respectively. Moreover and as shown in **Figure 2**, the band structure topologies are common and similar to all MMCs. Besides this, the dispersion of the bands nearest to the gap is nearly the same along the $\Gamma$-X and $\Gamma$-Y directions, while the electronic configuration of the metal plays a critical role in such dispersion. Furthermore, apart from the monolayer of GeSe, mono- and bi- layer GeS and bulk GeS exhibit indirect bandgap. Although, SnS has an indirect band gap (calculated with HSE) regardless of the layer number. The corresponding $E_g$ value is 1.96, 1.60, and 1.24 eV for monolayer, bilayer, and bulk SnS, respectively. In case of monolayer, bilayer, and bulk, the valence-band maxima (VBM) and conduction-band minima (CBM) are located along the $\Gamma$-X and $\Gamma$-Y lines (**Figure 2**). However, in case of monolayer, there are competing local CBM and VBM, which are very close in energy to the band edges. In a monolayer SnSe the direct gap of 1.44 eV is calculated along the $\Gamma$-X line. However, an additional direct transition at 1.60 eV (T1) occurs in the $\Gamma$-Y direction. In addition, the monolayer GeS shows an indirect bandgap of 2.32 eV, which is along the $\Gamma$-X and $\Gamma$-Y lines of CMB and VBM. Mono- and bi- layer GeSe shows a direct band gap of 1.54 eV (along $\Gamma$-X line) and 1.45 eV (along $\Gamma$-X line, near the X point), respectively. The energy band gap of such monolayers was verified by other methods, reported by Huang et al., Wang co-workers and Gong et al. (**Figure 2c**).[80-82] In this work the energy band alignment of CBM and VBM with respect to the vacuum level in MMCs with different number of layers has been estimated. The CBM exhibits a significant down shift, while the VBM remain relatively the same. Such change in the electronic band structure with layer number is quite similar among MMCs. The agreement was well fitted, except for the monolayer SnS, attributed to the presence of shallow core *d* bands. Apart from the mono and bi- layer MMCs, the bulk electronic structure is also shown in **Figure 2**. The calculated band gaps are 1.40, 1.00, 1.81, and 1.07 eV corresponding to the SnS, SnSe, GeS and GeSe respectively.

**Table 2.** *Summary of typical electronic properties of pristine MMC monolayers.*

| MMCs | Band transition | Band gap (eV) | VB (eV) | CB (eV) | References |
|------|----------------|---------------|---------|---------|------------|
| SiS (pristine) | Indirect | 1.44 | - | - | [79] |
|  |  | 1.37 | - | - | [64, 71] |
|  |  | (at the PBE level) |  |  |  |
| SiSe (pristine) | Indirect | 1.04 | - | - | [64] |
|  |  | (at the PBE level) |  |  |  |
| GeS | Indirect | 2.32 | 5.41 | 3.09 | [31, 80, 81] |
|  |  | 2.15 | 5.31 | 3.36 |  |
| GeSe | Direct | 1.54 | - | - | [31] |
|  |  | 1.59 | 4.82 | 3.23 | [80] |
| SnS | Indirect | 1.96 | - | - | [31] |
|  |  | 2.03 | 4.90 | 2.87 | [80] |
| SnSe | Direct | 1.44 |  |  | [31] |
|  | Indirect | 1.39 | 4.61 | 3.22 | [80] |

On the other hand, SiS and SiSe monolayer exhibit an indirect band gap of 1.44 eV and 1.04 eV, respectively.[64, 79] The MXs band structure has a significant anisotropy (top of the VBM) in the $\Gamma$-X and $\Gamma$-Y directions (**Figure 2b**). This leads to higher interlayer binding energy ($E_b$) value in MXs, which makes them relatively stable than the phosphorene (**Figure 2a**). Apart from the band structure, the spin-orbital coupling in monolayer MMCs has been estimated using a generalized gradient approximation (GAA) method. Such coupling is solely depends on the inversion symmetry of the crystal itself, which is related to the presence of odd/even number of layers in MXs. In general, MMCs monolayers exhibit a very similar electronic band structure with and without spin-orbital coupling. However, in case of SnSe (**Figure 2c**) and GeSe the spin orbital coupling is due to the conduction band splitting, while the VBM remains the same along $\Gamma$-X for all the cases. As shown in **Table 3**, the estimated spin orbital coupling in SnS (CBM) is the largest among MMCs. Moreover, it is striking that the spin-orbital splitting in MMCs conduction band exceeds that of other 2D materials, for which the corresponding splitting lies between 3 and 30 meV (**Table 3**).[83]

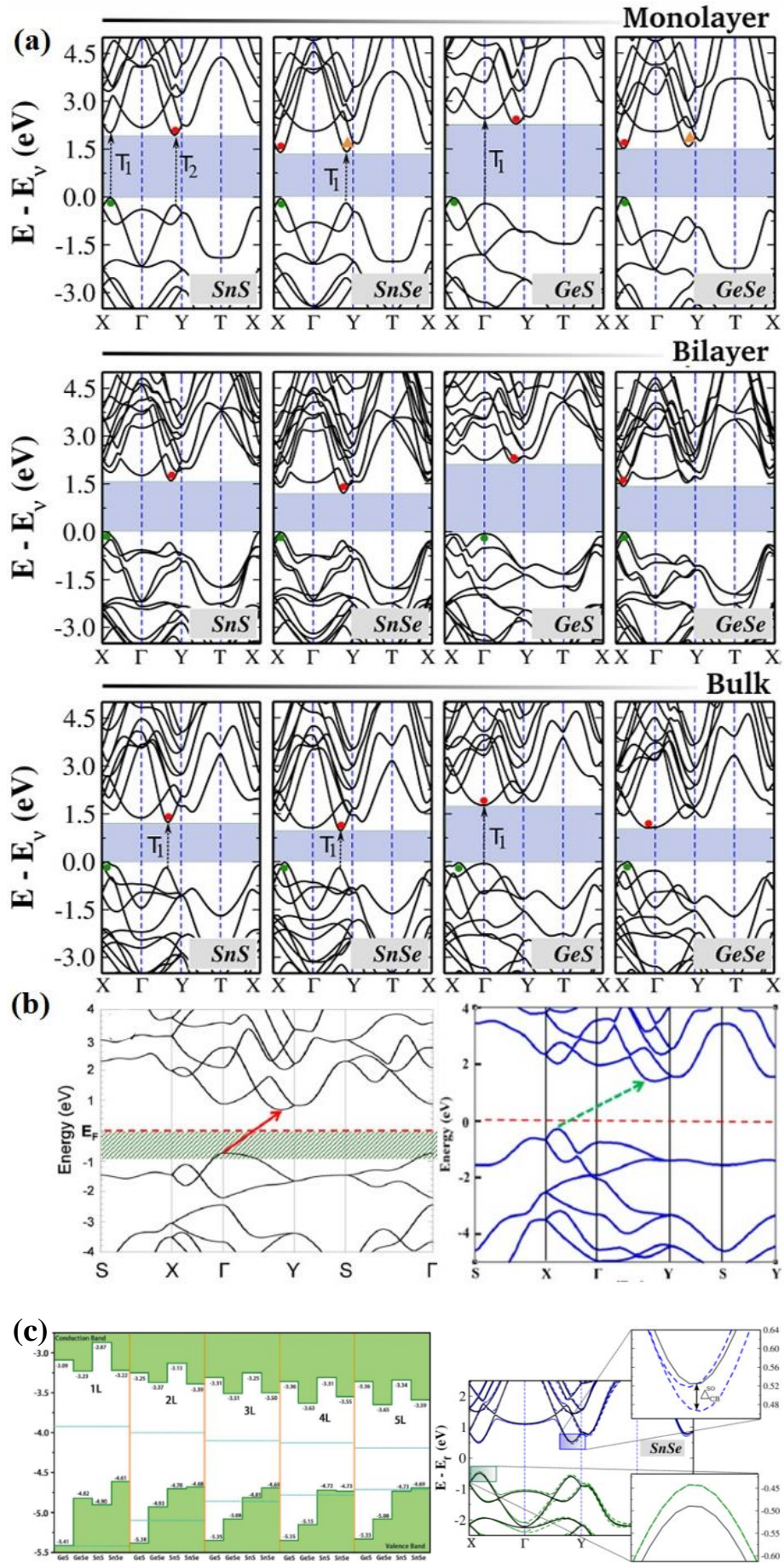

**Figure 2.** *a) Electronic band structures for monolayer, bilayer, and bulk MMCs (calculated using the Heyd-Scuseria-Ernzerhof (HSE06) hybrid functional). Solid circles are VBM and CBM. Black arrows (dashed) are possible direct transitions (T1 and T2) to points very close in energy to the VBM and CBM. Reproduced with permission.[31] Copyright 2015, American Physical Society. b) Electronic structure of: SiS monolayer, calculated using Density Functional Theory (DFT) (left panel) Reproduced with permission.[79] Copyright 2015, American Chemical Society.; SiSe monolayer calculated using HSE06 hybrid function (right panel). Reproduced with permission.[64] Copyright 2015, American Chemical Society. c) Left panel: Calculated band alignments of MXs of 1 to 5 layers (L) (at the HSE06 level) Reproduced with permission.[80] Copyright 2017, Royal Society of Chemistry. Right panel: Electronic band structure of SnSe monolayer with (dashed lines) and without (continuous lines) spin-orbit coupling effect. Reproduced with permission.[31] Copyright 2015, American Physical Society.*

The investigation of intrinsic electronic and optical properties of 2D materials is not only crucial for fundamental studies, but also from applications perspective. During the last decade, a series of reports have been focused on the theoretical investigation of such properties for monolayer 2D MMCs.[31, 33, 35, 39, 42, 80, 84-89] In particular Feng and co-workers[35] systematically investigated the electronic mobility in monolayer MMCs and adopted a phonon-limited scattering model to interpret the physical phenomena occurred. In a 2D system the mobility is provided by the formula $\mu_{2D} = \frac{e\hbar^3 C_{2D}}{k_b T m_e^* m_d (E_l^i)^2}$ where, $m_e^*$ is the effective mass of the electron along the transport direction and $m_d$ is the average carrier effective mass. Using on this formula, the predicted carrier mobilities for MMCs monolayers are in the order of $10^3$ to $10^5$ cm$^2$ V$^{-1}$ s$^{-1}$, while there are highly structural in-plane anisotropic in nature along armchair and zig-zag direction (**Figure 1**). For example, the electron mobilities of Ge and Sn based monochalcogenides are higher along the armchair and zig-zag direction, respectively. Exceptionally high carrier mobilities revealed in SnSe monolayer, attributed to the small effective masses and low deformation potential constants, along a particular direction (armchair or zig-zag). Such carrier mobilities are highly anisotropic compared to BP and other 2D TMD monolayers. Besides the high carrier mobilities, the monolayer MMCs exhibit piezoelectricity, with large piezoelectric constants.[90] Such piezoelectricity is more pronounced than that in other 2D systems, such as hexagonal BN and TMD monolayers.[42]

**Table 3.** *Optical and electronic properties of MMCs.*

| MMCs | Absorption coefficient (cm$^{-1}$) | Carrier mobility (cm$^2$V$^{-1}$s$^{-1}$) | Electronic properties | Spin orbital coupling | | References |
|---|---|---|---|---|---|---|
| | | | | $\Delta_{VB}^{SO}$ | $\Delta_{CB}^{SO}$ | |
| SiS | $10^5$-$10^7$ | $1.50\times10^3$ (zig-zag) $1.80\times10^4$ (armchair) | p-type | - | - | [62] |
| SiSe | - | - | - | - | - | - |
| GeS | $1.6\times10^5$ | $2.4\times10^3$ | p-type | 1 | 56 | [31, 36, 66, 87, 89] |
| GeSe | $10^5$ - $8\times10^5$ | $\sim10^3$ | p-type | 9 | 48 | [31, 35, 84] |
| SnS | $5\times10^4$ - $5.7\times10^5$ | $\sim10^3$ | | 8 | 87 | [31, 35, 89] |
| SnSe | $2\times10^4$-$9.5\times10^4$ | $10^4$ | p-type | 14 | 52 | [31, 35, 84, 88] |

## 3. Methods for MMCs preparation and characterizations

Since the first exfoliation of graphene, many methods have been developed to isolate ultrathin layers of 2D materials. All such methods can be divided into top-down and bottom-up ones. The top-down approach relies on the exfoliation of thin 2D crystals from their parent layered bulk crystals and the most important are based on mechanical (ME), liquid phase (LPE) and electrochemical exfoliation (EE).[4, 5, 91-93] While, the most important bottom up approaches include wet chemical synthesis (WCS), pulsed laser and chemical vapor deposition (PLD and CVD). Those methods are based on chemical reactions of certain precursors at given experimental conditions. In the following, we summarize the recent developments in the synthetic methods employed to produce ultrathin and single-layer MMCs.

## 3.1 Top-down methods

### 3.1.1 *Mechanical Exfoliation*

ME has been widely employed for obtaining few-layer and monolayers of graphene, TMDs and BP from their bulk counterparts.[1, 23, 94] Generally, the mechanical force is employed via scotch tape to weaken the van der Waals interaction between the adjacent layers of 2D bulk crystals and peel off single- or few-layered flakes. This technique can produce high-quality and clean surface crystals, which is favorable for both fundamental studies and technological applications. Besides this, the clean surface attained in mechanically exfoliated 2D materials makes them more suitable to stack and form good-quality van der Waals heterostructures.[95-97] To date, there are few reports on ME of thin layer of 2D MMCs, such as GeS,[98-100] GeSe,[101-103] SnS,[50, 57, 104] and SnSe.[36, 105, 106] Owing to the high interlayer binding energy of MMCs, the exfoliation of an atomically thin single layer is quite difficult.

Tan et al.[107] has isolated the ME of thin layers of GeS, with various thickness (65 nm to 8 nm). The crystal quality of the exfoliated flakes was verified with Raman spectroscopy. As mentioned above MMCs belong to the orthorhombic crystal structure Pnma ($D_{2h}^{16}$) crystal symmetry. In this structure, 24 phonon modes at the center of the Brillouin zone can be expressed as,

$$\Gamma = 4A_g + 2B_{1g} + 4B_{2g} + 2B_{3g} + 2A_u + 4B_{1u} + 2B_{2u} + 4B_{3u} \tag{1}$$

where, $A_g$, $B_{1g}$, $B_{2g}$, and $B_{3g}$ are optically active Raman modes.[108-110] The Raman spectra of GeS exhibited three characteristic peaks, which are assigned to $B_{3g}$, $A_g^1$, and $A_g^2$ phonon modes. The $B_{3g}$ and $A_g$ modes, in particular, correspond to the in-plane shear vibration of adjacent layers parallel to one another in the $x$ (armchair) and $y$ (zigzag) directions, respectively (**Figure 3c**). Furthermore, chemical vapor transport (CVT) grown single crystal bulk GeS was utilized for the mechanical exfoliation of thin nanosheets.[98] The isolated flakes were highly crystalline in nature, few tens of nanometer thick and micrometer-sized in lateral dimensions. Ulaganathan et al.[66] have isolated a ~28 nm-thin layer of GeS (**Figure 3a**), exhibiting a highly crystalline and large smooth surface, investigated for anisotropic electronic response. On the other hand, GeS flakes of variable thickness have been isolated by Li and co-workers.[98] AFM imaging of the multilayer GeS flakes showed a three-step stack of up to about few hundreds of nanometers. The thinnest flake obtained was measured to be ~ 40 nm (**Figure 3b**). In such thick flake four vibrational Raman modes were detected using unpolarized light, namely at 110, 210, 236, and 269 cm$^{-1}$ corresponding to $A_g$ and $B_{3g}$ symmetric modes; the position of such modes well agreed with theoretical predictions.[109] The anisotropic Raman response was clearly detected upon using parallel and perpendicular laser incident light.

Furthermore, few tens of nanometers thick GeSe flakes were micromechanically exfoliated by several groups.[103] In particular, Yang at al.[103] exfoliated 118.9 nm-thick flakes (**Figure 3d**), which were further investigated for their anisotropic optical properties. It is observed that the polarization dependent Raman spectra exhibited a periodic change in intensity with rotational angle (**Figure 3e**), corresponding to the two crystalline orientations along armchair and zigzag directions respectively. The respective spectra ware fitted using the classical Raman selection rules,[108, 109] expressed by the relation $I \infty |e_i.R.e_s|^2$, where, $e_i$ and $e_s$ are the unit polarization vectors of incident and scattered light, and $R$ is the Raman tensor.

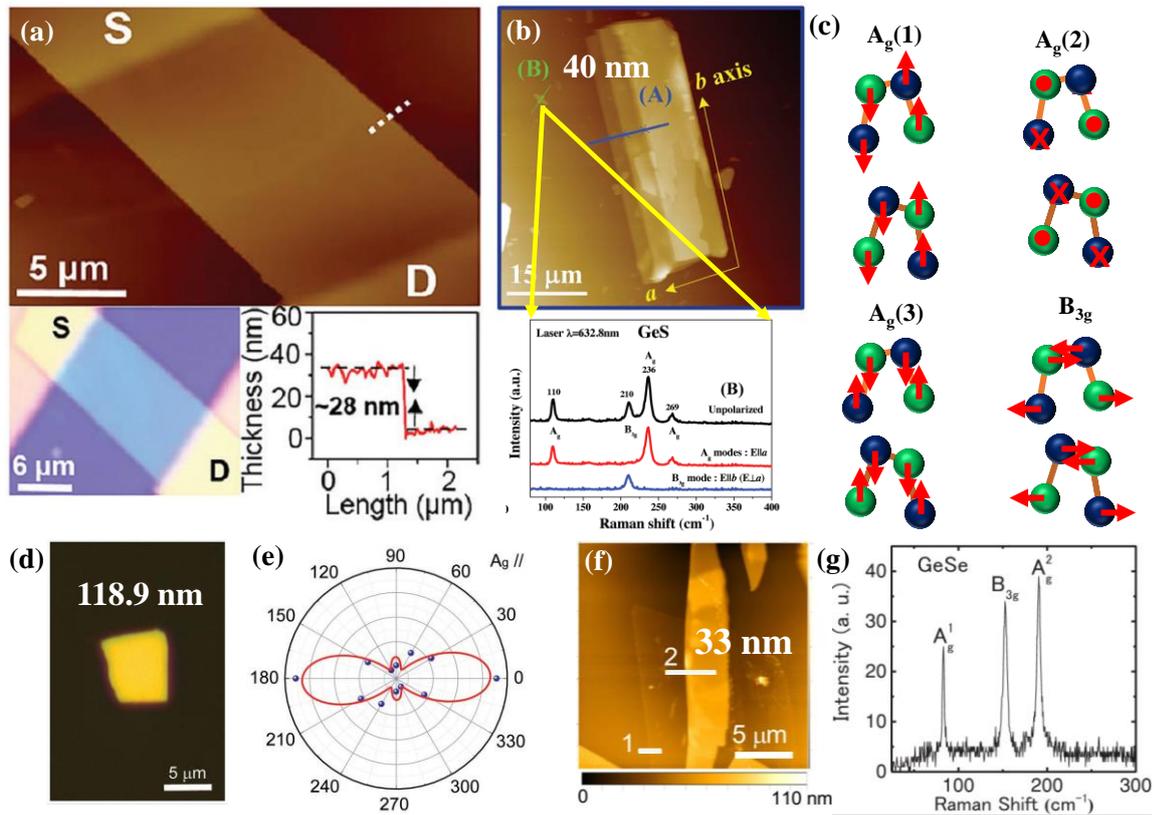

**Figure 3.** *Mechanically-exfoliated GeS and GeSe flakes. a) Atomic force (top panel) and optical microscopy (bottom panel) images. The thickness of the GeS flake is ∼28 nm (along the white dashed line (in top panel)). S and D correspond to source and drain respectively. Reproduced with permission.*[66] *Copyright 2016, Royal Society of Chemistry. b) AFM image of exfoliated GeS flake with a thickness from 40 nm (solid green line B) to three-step stack of up to about 270 nm (solid blue line A). Polarized Raman spectra of the 40 nm-thick GeS flake (bottom panel). Reproduced with permission.*[98] *Copyright 2017, Wiley-VCH. c) Schematic representation of the motion of atoms for the different Raman active modes in MXs. d) Optical image of GeSe flake, e) Angle resolved polar plot of the Raman peak intensity of $A_g$ mode (188*

Zhu and co-workers[102] isolated thinner GeSe flakes (230 to 14 nm) via micromechanical exfoliation to explore the anisotropic nature in current transport.[111, 112] The anisotropy in structure of a GeSe flake was realized by the angular resolved polarized Raman spectroscopy. Yang et al.[103] have adopted mechanical exfoliation to produce a few tens to hundreds nanometers –thick flakes, which are highly anisotropic in nature. In anisotropic materials, when the polarization direction of the incident laser light is parallel to a crystal orientation (armchair or zig-zag), the intensity of the optically active Raman modes reached a maximum or secondary maximum value.[113] A much thinner layers of GeSe flakes were isolated by Matsuda and co-workers[101] A stacked GeSe/MoS₂ heterojunction was reported (**Figure 3f**). A much thinner GeSe layer of 33 nm thickness was identified by the AFM height analyzer (**Figure 3f**). The Raman scattering spectrum of such thin layer revealed the characteristic peaks at 83.4, 152.5, and 190.4 cm⁻¹, assigned to $A_g^1$, $B_{3g}$ and $A_g^2$ phonon modes (**Figure 3g**), respectively.[111] The Raman peaks were shifted to the higher wavelength range compare to a 14 nm GeSe flake.[102] The observed phonon mode shift with lowering the layer number in GeSe is similar to other anisotropic 2D materials such as black phosphorus.[114-116] However, there is no systematic study of layer number dependent Raman scattering evolution in mechanically exfoliated GeSe flakes.

The successful isolation of few layers of SnS and SnSe flakes has been experimentally realized via micromechanical exfoliation.[36, 50, 57, 104, 106, 117] Nagashio and co-workers[118] produced thin layers of SnS flake by a micromechanical exfoliation method which was mediated by a tape exfoliation and aurum exfoliation (Au).The obtained flakes showed a wide range of distribution in size and thickness. The strong interlayer ionic bonding and the large electron distribution by lone pair electrons in Sn played a critical role to isolate them in large-area ultrathin layers. As also mentioned above the interlayer binding energy in a MXs is 146 meV per atom, which is much larger than those of graphite (24 meV per atom) and $MoS_2$ (38 meV per atom).[54, 56] However, the tape-exfoliated SnS flakes were of several micrometers in lateral dimensions, with tens of nm thickness. On the contrary, much larger SnS flakes were obtained via the Au-mediated exfoliation approach. In this method a strong semi-covalent bonding between Au and S atoms enabled the isolation of larger ultrathin SnS layers. A significant surface roughness was observed in exfoliated flakes (~0.1 nm), which may serve to produce a good quality interface with other 2D materials.[119-122] In another study, valley selective dichroism was identified in layered SnS by Chen et al.[50] The Raman vibrational modes, measured in a SnS flake about 109 nm-thick (**Figure 4a**), displayed a typical Lorentz shape peaking at 161, 191, and 217 $cm^{-1}$, corresponding to $B_{3g}$, $A_g^1$, and $A_g^2$ modes, respectively. As pointed out above, the structural anisotropy (along armchair and zig-zag direction) is an important feature in 2D MMCs, which originates from the stereo chemically active lone pair electrons in Sn 5s.[59, 60, 123] Therefore, the polarization dependence of the Raman scattering is a convenient way to determine the crystal orientation. Accordingly, the angular-resolved Raman scattering spectrum was recorded (**Figure 4b**) and the respective modes' intensity was displayed in a polar plot. The $A_g$ mode intensity well followed the $a\cos^2 2\theta + b$ dependence, where, $\theta$ is the polarization angle, $a$ and $b$ are fitting parameters. On the other hand the $B_{3g}$ mode intensity showed a $a\sin^2 2\theta + b$-dependence. Such observation of distinct fitting behavior of $A_g$, and $B_{3g}$ Raman modes attributed to the different Raman tensors governing them,[70, 124] which determine the structural anisotropy in armchair (θ=0º) and zigzag directions respectively. Furthermore, a much thinner layer was isolated by Nagashio and co-workers.[57]

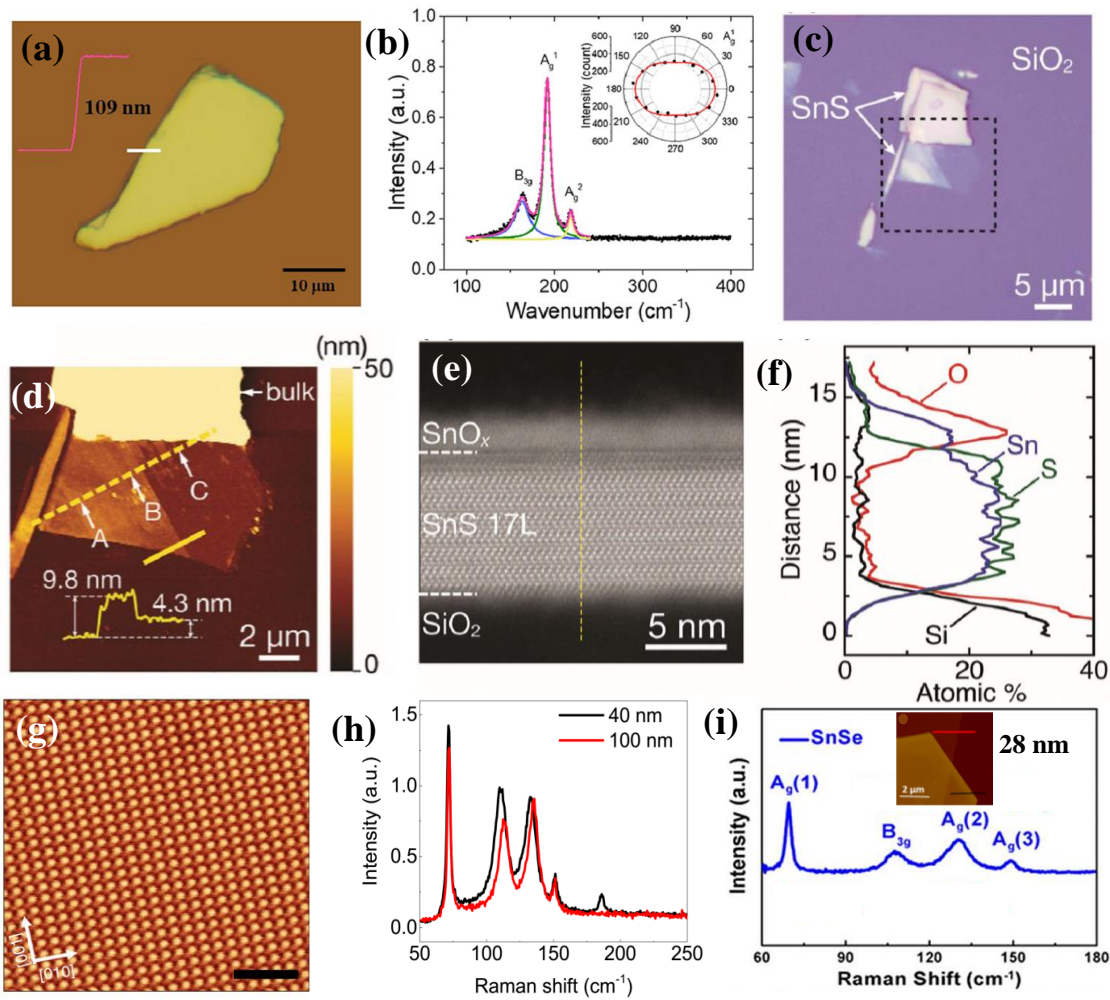

***Figure 4.*** a) *Optical micrograph of a mechanically exfoliated, 109 nm-thick, SnS flake on SiO₂/Si substrate; the corresponding AFM height profile is shown in the inset. b) Raman scattering spectra of the SnS flake. The solid lines (blue, green, and yellow) are Lorentz- fitting curves of the active Raman modes; The angular-resolved Raman intensity of the $A_g^1$ mode is shown in the inset. The θ=0º corresponds to the armchair, x direction of the crystal. Reproduced with permission,[50] Copyright 2018, American Chemical Society. c) Optical image of exfoliated SnS. d) AFM topography of the selected area in (c). Inset is the height profile along the solid yellow line. e) Cross-sectional high-angle annular dark-field scanning transmission electron microscopy (HAADF-STEM) image (at the point A in (d)). f) energy-*

*dispersive X-ray spectroscopy (EDS) depth profile along the dashed line in (e). Reproduced with permission,[57] Copyright 2018, Royal Society of Chemistry. g) Atomic-resolution non-contact atomic force microscope (nc-AFM) image of stoichiometric SnSe flake; the inset shows a typical optical image of SnSe₂ microdomains on SnSe. h) Raman spectra of the SnSe flake shown in g) Reproduced from[106] Copyright 2018 Author(s), licensed under a Creative Commons Attribution 4.0 License and i) Room-temperature Raman spectra of SnSe flake. Inset shows AFM height profile along the red line. Reproduced with permission,[125] Copyright 2017, American Chemical Society.*

These cotch tape and Au-exfoliation methods were adopted to isolate such thin flakes. The Au- exfoliated SnS flakes exhibited various thicknesses (**Figure 4c, d**), with the thinnest layer to be ~4.3 nm. However, upon Au exfoliation, surface modification was observed, which badly affects the nanosheet properties. The formation of amorphous tin oxide (a-$SnO_x$) is revealed at the surface of the intrinsic SnS, in particular (**Figure 4e, f**).

Cho et al.[105] exfoliated of a single crystal, prepared by the modified Bridgman technique to produce single-phase orthorhombic SnSe flakes, with the Pnma space group. Similar methodology was employed by Yang et. al.[36] to prepare bulk SnSe flakes ranging from few tens to hundreds of nanometer thickness. It is shown that a 71 nm-thick SnSe crystal exhibited orthogonal geometry. In such crystal the intersection angle of the two crystal planes is determined to be 90º.[126] A strong 4-fold anisotropy with a period of 90º and minimum intensities along the 0° and 90° directions, identified as zig-zag and armchair directions respectively. Such anisotropic Raman spectral behavior in MXs is strongly dependent on the phonon symmetry, which was observed in other 2D materials having in-plane structural anisotropy.[127-131] In another study, a much thinner flake of SnSe was isolated by Zheng and co-workers,[106] The high quality of the SnSe flake was visualized by non-contact atomic force microscopy (nc-AFM) showing that very few vacancies were present. The crystal phase of the SnSe flake was identified by the corresponding Raman spectra (**Figure 4g, h**). Interestingly, Wei and co-workers[125] have isolated a 28 nm-thick SnSe flake (**Figure 4i**) to produce a heterostructure with molybdenum disulfide (MoS$_2$). The heterostructure was further used for electronic and optoelectronic investigations. In particular, the vibrational modes exhibited four characteristic peaks at 69.5, 109.0, 130.6, and 149.5 cm$^{-1}$, associated to A$_g$ and B$_{3g}$ active Raman modes.

In spite of the great advantages of the ME method, including that it is a relatively easy process and provides high-crystalline quality 2D flakes, the MMCs produced suffer from size repeatability and reproducibility of the layer number. At the same time, owing to the strong interlayer force and in-plane anisotropy the isolation of a single layer is still a challenge. In contrast, liquid phase exfoliation and chemical vapor deposition have been employed for the isolation of thin layer, large area, and controllable in morphology nanosheets.

*3.1.2 Solution or liquid phase exfoliation*

Liquid phase exfoliation (LPE) is a well-known strategy to produce ultrathin layered 2D materials and has thus been extensively used to exfoliate 2D layered graphene and other crystals.[4, 91, 132-134] The method has been developed and advanced by the Coleman and co-workers[132, 135] in 2008 and has considered to be one of the most promising and simplest routes for the production of 2D materials in a large scale. It generally involves the process in which bulk layered crystals or powders are dispersed in a suitable solvent and are subjected to ultrasonication for a certain amount of time. Following ultrasonication, the suspension is centrifuges to separate the unexfoliated bulk and exfoliated thin nanosheets (NSs). The fundamental idea is that ultrasonic waves can induce liquid cavitation in the dispersion, which in-turn leads to cavitation bubbles. The generated bubbles collapse onto the dispersed material. As a consequence, intensive tensile stress is applied to the dispersed bulk crystals, giving rise to thin layers exfoliation.

A critical factor for the success of LPE of layered 2D materials is the selection of appropriate solvent. Indeed, the solvent screening critically depends on the Hildebrand and Hansen solubility parameters.[136, 137] However, recent reports have demonstrated that neither Hildebrand nor the Hansen parameters can fully describe the solvent-solute interaction during the LPE of 2D materials.[138] Thermodynamic laws suggest that a lower difference in surface energy between the 2D material to be exfoliated and the solvent is beneficial for LPE.[132, 137] Beside this, the ultrasonication time, the temperature during ultrasonication, as well as the centrifugation rate, are critical parameters for efficient exfoliation, both from qualitative and quantitative point of view.[135, 139] To date, few reports on the LPE synthesis of ultrathin layer MMCs have been presented, referring to GeS, GeSe, SnS and SnSe (**Table 4**).[34, 110, 140-147] In contrast, there is no experimental evidence on the LPE of SiS and SiSe.

**Table 4**. *Summary of the experimental synthesis of metal monochalcogenides (GeS, GeSe, SnS and SnSe) via LPE.*

| MMCs | Synthesis Method | Solvent | Average thickness (nm) | Lateral dimension (nm) | Layer number | References |
|---|---|---|---|---|---|---|
| GeS | LPE | Hexene, ethanol, IPA, NMP, DMF, acetone, chloroform | 2.87±0.65 (NMP) | 66.91±0.42 | 6 layers | [140] |
| GeS | LPE | NMP | 1.3±0.1 | - | Tri-layer | [143] |

| GeSe | LPE | Ethanol | 2 | 50-200 | 4 layers | [141] |
|------|-----|---------|---|--------|----------|-------|
| GeSe | LPE | NMP | 4.3 ± 0.2 | - | 8-9 layers | [142] |
| SnS | LPE | NMP | 4.2±0.24 | 5-100 | 6-8 layers | [34] |
| SnS | LPE | NMP | 6 | - | Few layers | [144] |
| SnS | LPE | IPA | 4-8 | 50 | Few layers | [146] |
| SnS | LPE | DMF | 4.5 | 400-900 | 8 layers | [147] |
| SnS | LPE | Acetone | 1.10 | 170 | Bilayer | [110] |
| SnSe | LPE | IPA | 4.3 | 50-500 | 7 layers | [148] |
| SnSe | LPE | NMP | 2.5 | 50-200 | 4 layers | [149] |

Hersam and co-workers[140] isolated few layer GeS NSs via anhydrous solvents LPE of bulk powder (**Figure 5a**). Considering the solvent and solute chemical properties a series of organic solvents exhibiting different surface tensions were investigated. It is reported that the isolated sheets in NMP showed the darkest dispersion (**Figure 5b, c**), which is an evidence of this solvent suitability to efficiently stabilizing the exfoliated GeS sheets. The structural integrity of exfoliated crystals was verified by studying the lattice vibration modes (**Figure 5d**), including the $B_{3g}$ mode (213 cm$^{-1}$), corresponding to the in-plane shear vibration of parallel layers in the zig-zag direction and the $A_g$ modes (112, 240, and 270 cm$^{-1}$) corresponding to shear vibration of parallel layers in the armchair direction (a-axis). A standard probability density function was used to fit the characteristic size histograms (through AFM) recorded for the exfoliated flakes (**Figure 5e-g**), namely

$$y = y_0 + \frac{A}{x\sigma\sqrt{2\pi}} e^{-\frac{\left(\ln\frac{x}{\mu}\right)^2}{(2\sigma^2)}}$$

(2)

where, $y_0$ is a constant offset, A is a constant prefactor, $x$ is either the flake thickness or length, $\sigma$ is log normal standard deviation, and $\mu$ is the long normal mean. In another study, a cascade centrifugation method was employed to achieve much thinner layer of GeS sheets by Fan et al.[143] The LPE method was adopted to exfoliate bulk powders in NMP solution in which different centrifuge settings were employed to realize thinner layers. The thickness was found to be as thin as 13.2±2 nm, 4.2±0.3 nm, 3.2±0.2 nm, and 1.3±0.1 nm, corresponding to 2-4k, 4-6k, 6-8k, and 8-10k, settings respectively (**Figure 5h**). The thinnest layer of GeS achieved was 1.3±0.1 nm (8-10k), which corresponds to approximately two monolayers.

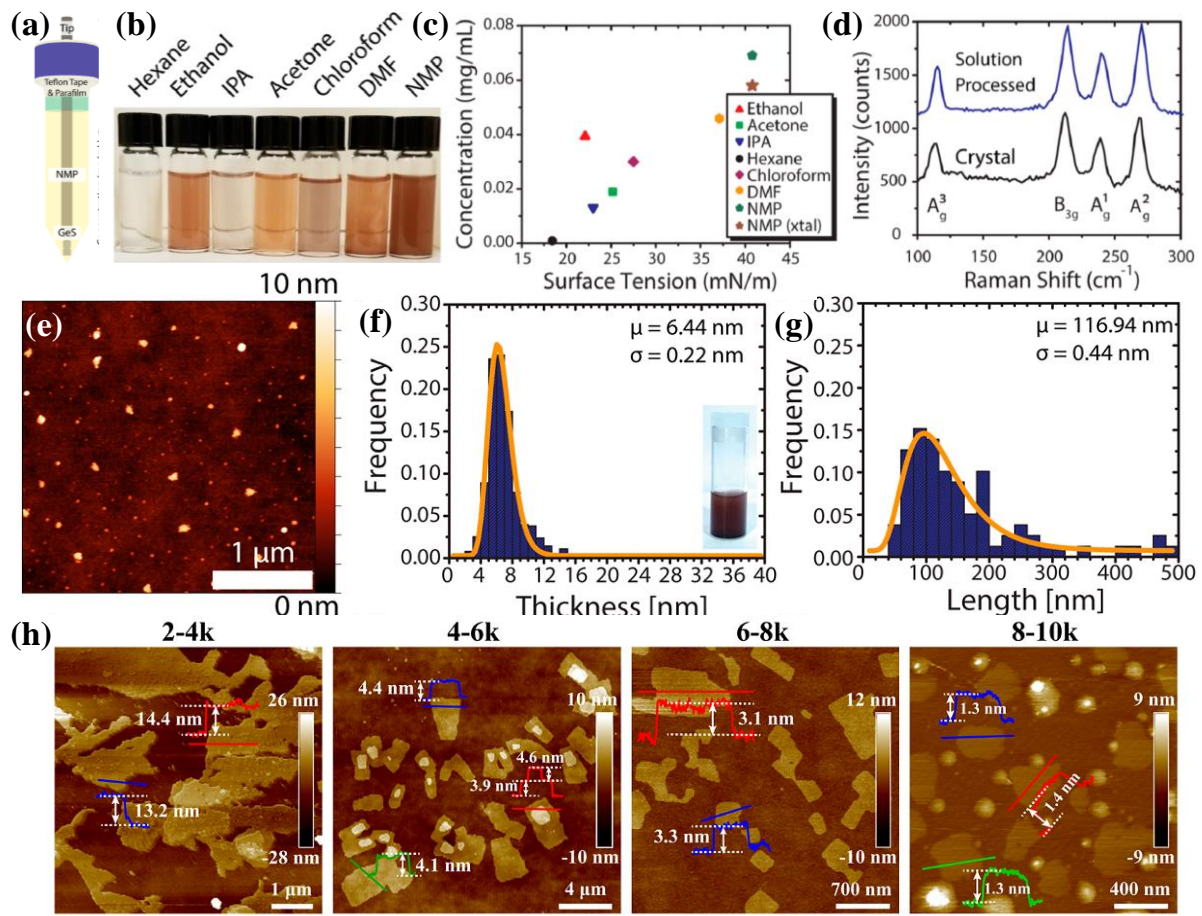

**Figure 5.** *a) Schematic of Tip Sonicator. The solution tube is sealed with Teflon tape and parafilm to minimize exposure to ambient atmosphere. b) Exfoliated GeS NSs in various solvents (centrifuged at 1000 rpm for 10 min), c) Concentration of exfoliated bulk powder of*

*GeS as a function of solvent surface tension. The NMP exfoliated GeS is considered as a reference concentration. d) Raman spectra of a bulk GeS crystal and solution-processed GeS flakes. e) AFM topography image of NSs. f) TEM image of NSs. g) Thickness (inset: as prepared GeS dispersion) and h) lateral size histograms (from statistical TEM analysis) of the as-exfoliated GeS NSs. i) Thickness (inset: centrifuged GeS dispersion at 500 rpm) and j) lateral size histograms (from statistical TEM analysis) of the centrifuged GeS NSs. Reproduced with permission[140] 2018, American chemical society. h) AFM images, corresponding height profiles, and thickness distribution histograms of LPE GeS NSs collected with different centrifugation speeds. Reproduced with permission.[143] Copyright 2019, American chemical society.*

A different method was adopted by Ye et al.[141] The GeSe flakes have prepared via sonication-assisted LPE with cascaded centrifugation. A series of solvents were tested, namely, IPA, DMF, CHP, SDS, and SC, leading to GeSe NSs with quite different lateral size and thickness, which are redispersible. The CHP exfoliated NSs are produced largest lateral size, while exfoliation in IPA produces much smaller sizes. However, the thinner layer, achieved in ethanol, showed lateral dimensions in the range 50-200 nm, which is highly single crystalline in nature with hexagonal structures (**Figure 6a, b**). Even thinner layers were subsequently obtained upon centrifugation at different conditions. In particular, the thickness of the GeSe NSs, obtained at the centrifugation speeds of 9k, 6k and 3k rpm, showed an average thickness of 2, 5.5 and 6 nm, respectively (**Figure 6c-h**). The orthorhombic GeSe crystals obtained belong to the $D_{2h}^{16}$ symmetry, which exhibits 12 optically active Raman modes ($4A_g+2B_{1g}+4B_{2g}+2B_{3g}$).[108, 111] Three such modes which are optically active and peaked at 80, 150 and 180 cm$^{-1}$, corresponding to $A_g^3$, $B_{3g}^1$, and $A_g^1$ respectively. Beyond this time, the isolated GeSe sheets showed a band gap variation from their bulk counterpart. Zhang co-workers[142] have systematically investigated the sonication effects during the LPE of GeSe NSs. The typical process is demonstrated in **Figure 6i**. A series of organic solvents were tested including NMP, DMF, water, ethanol and IPA, and the effect of centrifugation speed (within the ranges 1-14 k, 2-14 k, and 3-14 k) on thickness and lateral size of the exfoliated GeSe was examined.

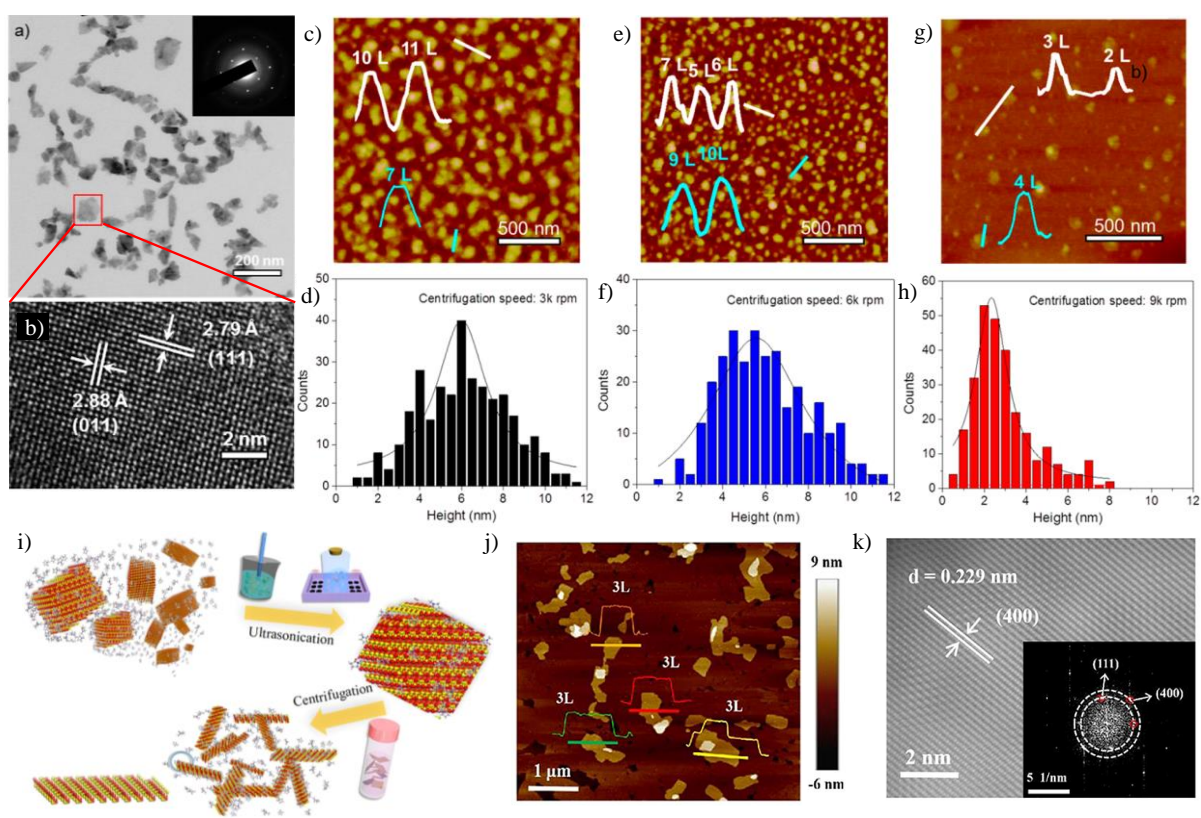

**Figure 6.** *TEM image of a) GeSe sheets (Inset: Selected area electron diffraction (SAED) pattern taken from a single flake marked by red rectangle), b) High resolution transmission electron microscopy (HRTEM) image of a GeSe sheet. Height-mode AFM images of exfoliated GeSe NSs collected at different centrifugation speed of c) 3k, e) 6k, and g) 9k respectively (Inset: height profiles). d, f, h) Histogram analysis for the thickness of GeSe NSs such as those shown in the panels c, e, g, respectively. Reproduced with permission*[141] *Copyright 2017, American chemical society. i) Schematic illustration of the LPE process used to obtain ultrathin GeSe NSs. j) AFM topography image and height profile of LPE GeSe NSs. and k) HRTEM image of exfoliated GeSe NSs in NMP solvent; the inset shows the SAED pattern. Reproduced with permission*[142] *2019, American chemical society.*

A significantly decreased size distribution and average lateral size with no discernable oxide phase impurities were detected in the NMP-exfoliated GeSe flakes. While, the lowest average thickness was estimated to be around 4.3 ± 0.2 nm (**Figure 6j**), which correspond to tri-layers based on the previous theoretical calculations[150-152] Beside this, the exfoliated GeSe sheets are highly crystalline with clear crystal lattice spacing (≈0.23 nm) (**Figure 6k**).

In 2015, O'Brien and co-workers[34] first reported the LPE of anisotropic layered SnS. Following this work, many research groups have been intensively focused on exfoliation of ultrathin SnS layers.[110, 144, 145] In the first report,[34] bulk SnS powder was dissolved in NMP and ultrasonication was employed to induce cavitation effects in the dispersion. In particular, the isolated SnS sheets (1500 rpm (Sol A) and 10000 rpm (Sol B)) exhibited an average height of 4.1 nm and 7.8 nm, respectively (**Figure 7a**), which corresponding to multilayer SnS. On the contrary, the exfoliated NSs were 50-100 nm in lateral dimension (**Figure 7c**), often with aspect ratios of around 1.5-2.0. The sharpening in the Raman modes ($A_g$, $B_{3g}$, and $B_{3u}$) in SnS NSs compared to the broad bulk ones was another examination of the thin nature of the exfoliated flakes (**Figure 7b**). In another approach, both a bath and tip sonicator have been employed to create the cavitation events in SnS colloidal dispersion[144] Liquid cascade centrifugation (LCC) was utilized to prepare size-selected 2D SnS NSs in NMP The thinnest sheets of 6.0 nm, comprising 10 or more monolayers, were seperated. The isolation of thin layer SnS using LCC method is an effective way to exfoliate few layer SnS sheets, which is in good agreement with other 2D materials prepared by the LCC technique.[153, 154] The size selective Raman modes appeared gradually blue-shifted as the thickness of the 2D NSs decreases, which is attributed to the confined oscillation when more layers of SnS are added and bonded by van der Waals forces. Furthermore, a lower boiling point solvent, isopropyl alcohol (IPA), was tested as an exfoliation medium by Zhang and co-workers[145] A similar approach of LPE by the O'Brien et al.[34] works has followed. The isolated SnS flakes showed a 50 nm lateral dimension with average thickness of 6 nm. A high quality crystal with lattice spacing was identified, which confirmed the crystal phase integrity, even after exfoliation. The overall investigation indicated that the low boiling point solvent is more suitable to isolate thin layer SnS. However, the exfoliated SnS sheets are limited to few nanometers-thick.

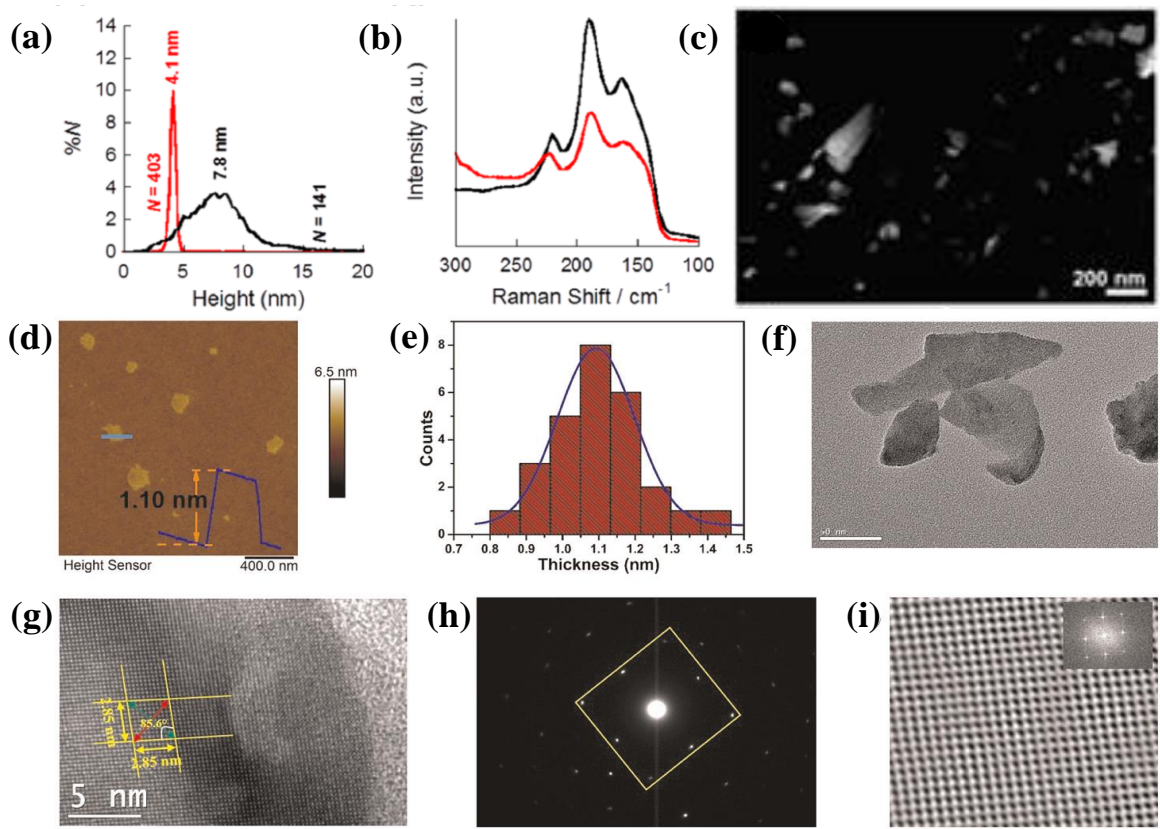

**Figure 7.** *Liquid phase exfoliated SnS nanosheets: a) Particle size distribution for NS height, as determined for SnS sol-A (black curve) and SnS sol-B (red curve); b) Raman scattering spectra of SnS samples; c) Low resolution High-angle annular dark-field (HAADF) scanning transmission electron microscopy (STEM) image of SnS; f) Atomic resolution HAADF STEM image of a SnS NS. Reproduced with permission,[34] Copyright 2015, from American chemical society. As-synthesized ultrathin SnS layers: d) Atomic force microscopy image (scale bar 400 nm); e) Histograms of thickness distribution; f) Low resolution TEM image (scale bar is 50 nm); g) HRTEM image (scale bar is 5 nm), The length of lattice fringes (10 fringes) was measured to be 2.85 nm along both different directions , h) SAED pattern; i) Fast Fourier*

*Transform filtered atomic resolution of the selected area. Inset: FFT pattern of the selected region in HRTEM image. Adapted with permission.[110] Copyright 2020 Author(s), licensed under a Creative Commons Attribution 4.0 License, Nature Publishing Journal.*

Most recently, Sarkar et al.[110] has reported the isolation of thinner SnS sheets via pushing one step forward the O'Brien's[34] method. An organic solvent was chosen to exfoliate bulk SnS crystal in a bath sonicator with a power of 100 W and 40 kHz frequency for 10 h. Acetone, in particular, was identified as the most suitable solvent for the successful exfoliation of electronic grade ultrathin SnS layers. The separated SnS sheets showed an average thickness of ~1.10 nm and average lateral dimensions of ~170 nm (**Figure 7d-f**). Those sheets are highly crystalline in nature, as perfect rhombus-like lattice fringes with sharp SAED patterns were observed (**Figure 7g-i**). Besides this, the optical properties of SnS sheets were explored towards thermoelectric and nanophotonic applications.

Nan and co-workers[148] has utilized the LPE of SnSe in a low boiling point solvent IPA. The thinnest NS (~1 nm), comprising two monolayers of SnSe, was obtained.[155] Very recently, Ye et al.[149] have investigated the intensive LPE of SnSe in seven different solvents and various centrifugation conditions (cascade centrifugation). Among the solvents tested, the NMP-exfoliated SnSe exhibited the larger, in lateral dimensions, flakes. A cascade centrifugation was employed, during which the NSs, collected at 8k rpm exhibited lateral sizes in the range ~50-200 nm (**Figure 8a, b**). The obtained average thickness of NSs 9.5, 6, and 2.5 nm, corresponding to layer numbers of 16, 10, 4 respectively (**Figure 8g-i**). The orthorhombic diffraction pattern, together with the lattice spacing and the highly crystalline order are in well agreement with the crystal structure (Pnma) of SnSe (**Figure 8c**). while, the relative presence of Sn and Se confirm the preservation of elemental composition (**Figure 8d-f**). The collected Raman spectra at different centrifugation speed exhibited four characteristics peaks. Considering that SnSe belongs to Pnma ($D_{2h}^{16}$) symmetry and thus have 12 active Raman modes, the peaks appeared at 70, 105, 127 and 150 cm$^{-1}$, correspond to the $A_g^1$, $A_g^2$, $B_{3g}^1$, and $A_g^2$, respectively (**Figur 8j**). Besides this, the chemical composition (molar ratio of Sn:Se) of exfoliated SnSe was determined to be 1:1. On the contrary, the high-resolution X-ray photoelectron spectroscopy (XPS) spectral investigation was revealed Sn core level Sn 3d$_{3/2}$and Sn 3d$_{5/2}$, while Se shown Se 3d$_{3/2}$and Se 3d$_{5/2}$ doublets (**Figure 8k, l**). Accordingly, an electronic grade ultrathin layer of SnSe has not yet been realized.

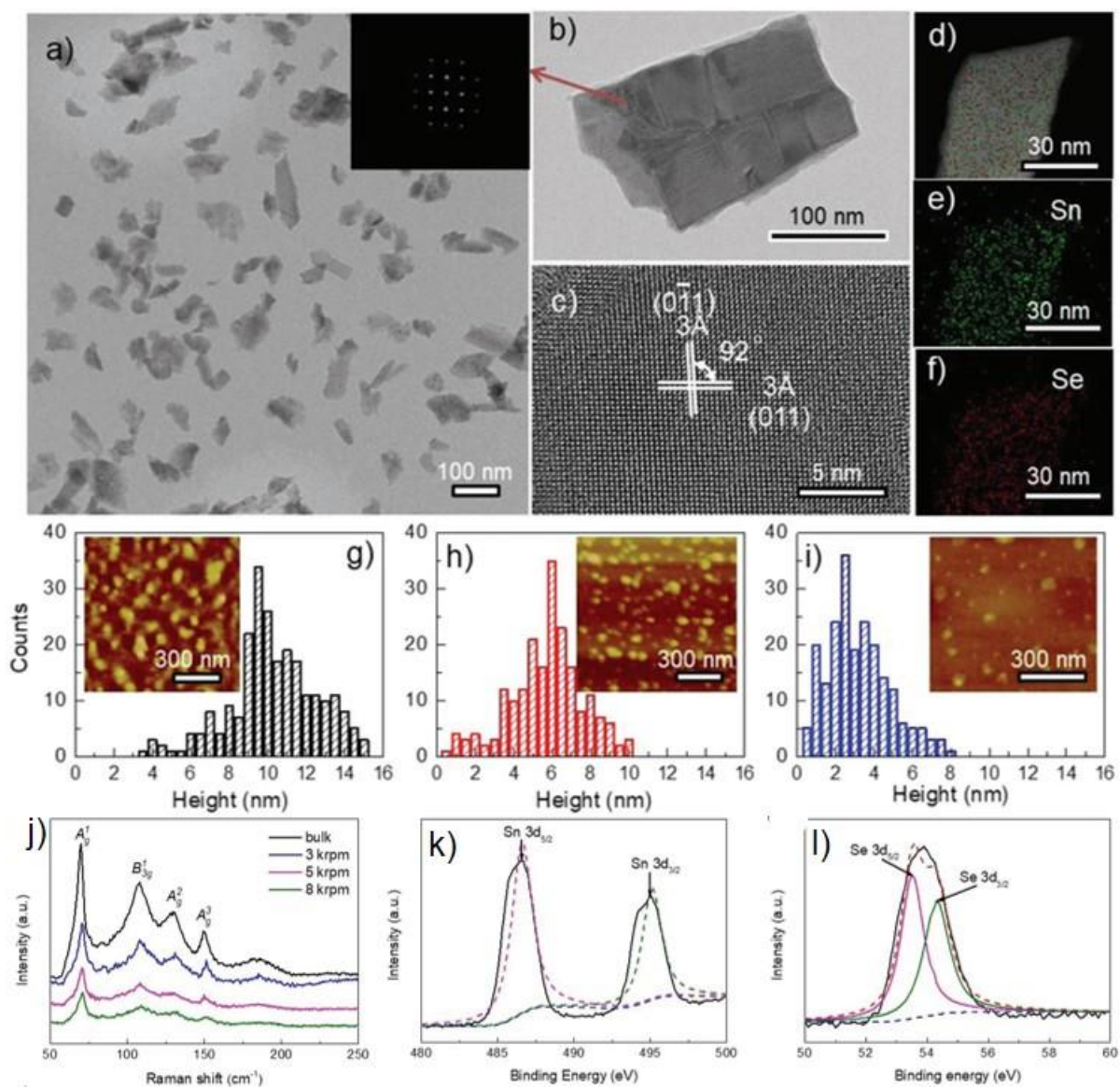

**Figure 8.** *a) TEM image of exfoliated SnSe NSs; b) a single SnSe flake (Inset: FFT image of the SnSe NS in (b)); c) HRTEM image of SnSe NSs; d-f) STEM image and the elemental mapping of Sn and Se; g-i) Histogram of thickness distribution and inset shows corresponding*

*AFM images of exfoliated SnSe NSs collected at various centrifugation speeds; j) Raman spectra of bulk SnSe and exfoliated NSs; k), l) High resolution XPS spectra of core level of Sn3d and Se 3d region. Reproduced with permission.[149] Copyright 2019, Wiley-VCH.*

Most interestingly, it is found that the low boiling point solvents, which have considerably lower surface tension, are suitable for the exfoliation of thin layer MMCs, particularly SnS and SnSe. The surface energy of such solvents played crucial role to create cavitation during ultrasonication process.

## 3.2 Bottom up methods via Vapor Phase Deposition

The vapor phase deposition process is widely used for the synthesis of atomic scale 2D semiconducting layered materials[6, 156-158] and is divided into the physical (PVD) and chemical vapor deposition (CVD) methods. In the PVD process no chemical reactions are taking place. In case of 2D MMCs' growth, either a single solid precursor of the final product is used, or co-deposition of the chalcogen and transition metal precursors on a substrate takes place. On the other hand, the CVD process employs reactive precursors and high vacuum, in which the precursors react and/or decompose on the surface of the substrate at high temperature to form ultrathin flakes. This process is one of the most effective method to realize large area growth of atomically thin layers of 2D TMDs. Below we summarize the most important reports to synthesize 2D MMCs, such as GeS,[159] GeSe,[68, 160] SnS,[70, 124, 161, 162] and SnSe[61, 163-165], via vapor phase deposition.

Sutter and co-workers[159] synthesized a few layer GeS flake via vapor transport. For this purpose, the GeS powder was heated to temperatures in the range of 430-450°C, while the mica substrate temperature was varied from 320 to 350ºC. The substrate temperature was played a significant role on the dimensionality of GeS flakes attained (**Figure 9a-e**). In particular, the use of low substrate temperature led to small-sized NSs (edge length of ~1.5 μm) (**Figure 9f**).

Flakes of larger size were obtained at a higher substrate temperature of 340ºC (**Figure 9g**). A bimodal size distribution was particularly observed in that case due to the secondary nucleation of the smaller flakes. The thickness of as-grown GeS varied from 77 to 24 nm (**Figure 9h-j**). However, the average flake thickness, identified by AFM height was found to be 34.5 nm. The crystal structure and morphology of the flakes were investigated by TEM. A perfect orthorhombic crystal structure with faceted shapes and clear crystal planes were revealed in GeS (**Figure 9l, m**). Those GeS flakes are single-crystalline with orthorhombic space group *Pnma* (**Figure 9l-n**). The vibrational properties of as grown GeS flake (340ºC) was investigated by polarized Raman spectroscopy (**Figure 9o**), showing the characteristic modes of vibrations ($A_g$ and $B_{3g}$). It is also observed that the intensity of the phonon modes in Raman spectra was changing with incident light polarization.

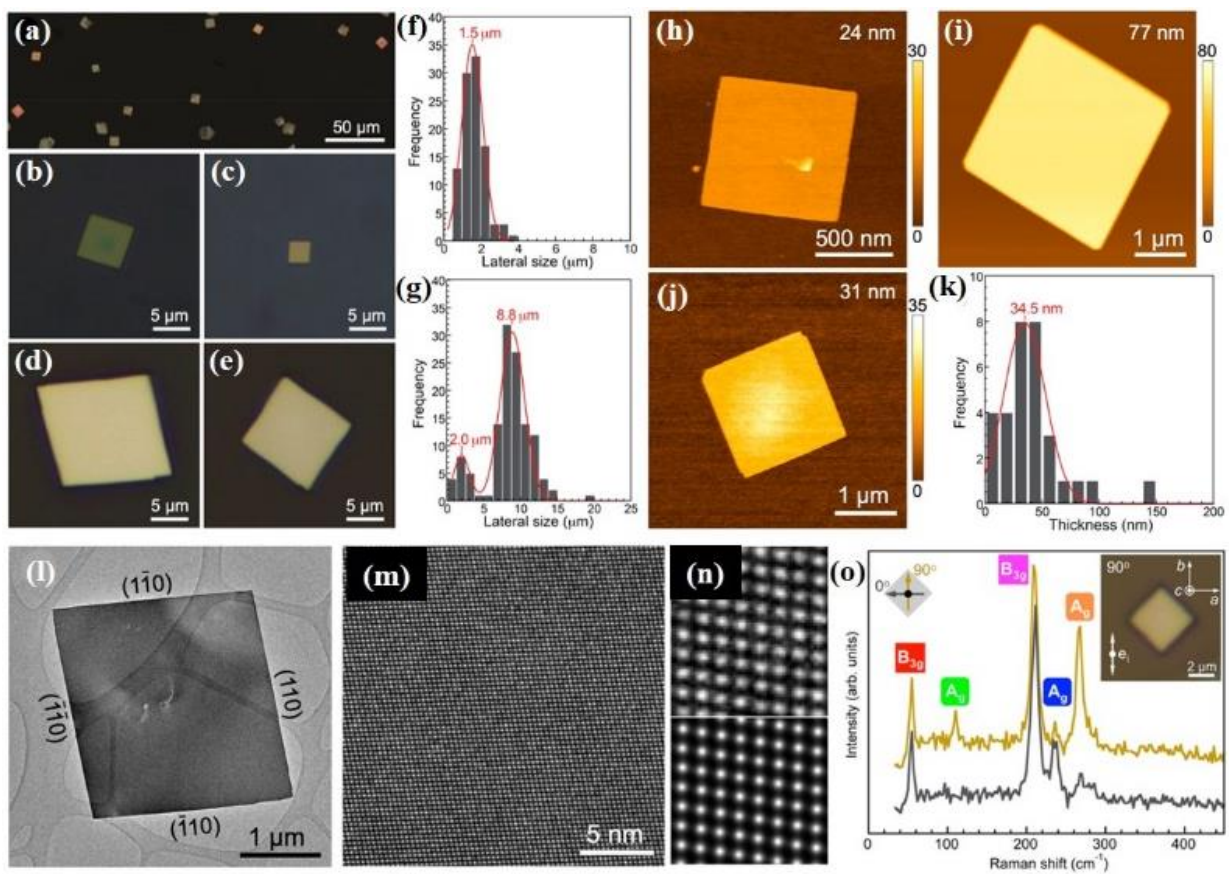

**Figure 9.** *a) Optical image of as grown GeS flakes on mica substrate. (b-e) Higher magnification optical images. Flake size distribution obtained by growth f) Monomodal at*

*320°C g) Bimodal at 340°C (growth time 10 min). h-j) AFM topography image of GeS flakes with different thickness. K) Thickness distribution of GeS flakes grown at 340 °C for 10 min. l, m) TEM images of low and high-resolution SnS flakes. n) Higher magnification view of (m) (top), and multislice image simulation (bottom). o) Anisotropic Raman spectra of synthesized GeS flake (thickness ∼500 nm) obtained with 0°and 90°orientations of the incident laser light linear polarization. Laser excitation wavelength was 532 nm with 16.8 µW laser incident power. Inset: an optical image of the measured flake and the direction of the incident electric field vector. Reproduced with permission.[159] Copyright 2019, American chemical society.*

Mukherjee et al.[68] has reported the growth of high-quality, single-crystalline, micrometer-sized 2D GeSe NSs using a CVD process with various substrate temperatures. A horizontal single-zone tube furnace was used for the synthesis, in which a mixture of Ge and Se powders (1:1 molar ratio) was heated at 480ºC for 4h under high mTorr vacuum. The dynamical behavior of such CVD process is represented in **Figure 10a**. The process involved the sublimation of bulk GeSe powder (source) into gaseous products and their subsequent transport be the carrier gas (Ar) followed by condensation of the gas molecules onto the substrate, placed at the temperature range 390-430ºC. The constant flow of the sublimated gas molecules and their subsequent adsorption and recrystallization to growth sites assisted the growth of the nanostructures. The obtained nanostructures were found to exhibit different morphologies (**Figure 10b-d**), depending on the substrate temperate (zones 1, 2, 3). The growth of different crystalline nanostructures strongly depends on the concentration gradient of GeSe vapor at the lower temperature zone (zone 3, large distance from the source). The typical lateral and longitudinal dimension of the NSs are measured to be in the range of 4-160 µm and 60-140 nm, respectively (**Figure 10e-f**).

Furthermore, Hu et al.[166] have reported a high quality single-crystalline ultrathin layer of 2D GeSe flakes, which are synthesized by a salt-assisted CVD method (**Figure 10g, h**). In this process, mixed $GeSe_2$ and KCl powder was heated to 550°C for 0.5h. During the heating process a mixed gas of $H_2$ and Ar was passed through the quartz tube used in the process. The $GeSe_2$ was easily reduced to GeSe with $H_2$ ($GeSe_2+H_2 \rightarrow GeSe + H_2Se$); in this reaction the salt KCl used played an assisting role. It was observed that the as grown thinnest GeSe flakes on mica substrates was ~5 nm (**Figure 10i**), corresponding to 8 monolayers. The Raman spectra of GeSe flakes with different thickness (5, 9, 15 nm) exhibited four characteristic modes of vibration, $A_g^1$, $B_{3g}$, $A_g^2$, and $A_g^3$, respectively (**Figure 10j**), peaked at 81.8, 151.9, 173.5, and 189.2 $cm^{-1}$ for a 15 nm-thick flake. These Raman modes became blue- shifted upon increasing the layer number. Such blue shift can be attributed to the strong interlayer coupling in thicker GeSe NSs. Similar shifts were observed in many 2D materials, including graphene, $MoS_2$, and GaSe.[166-169]

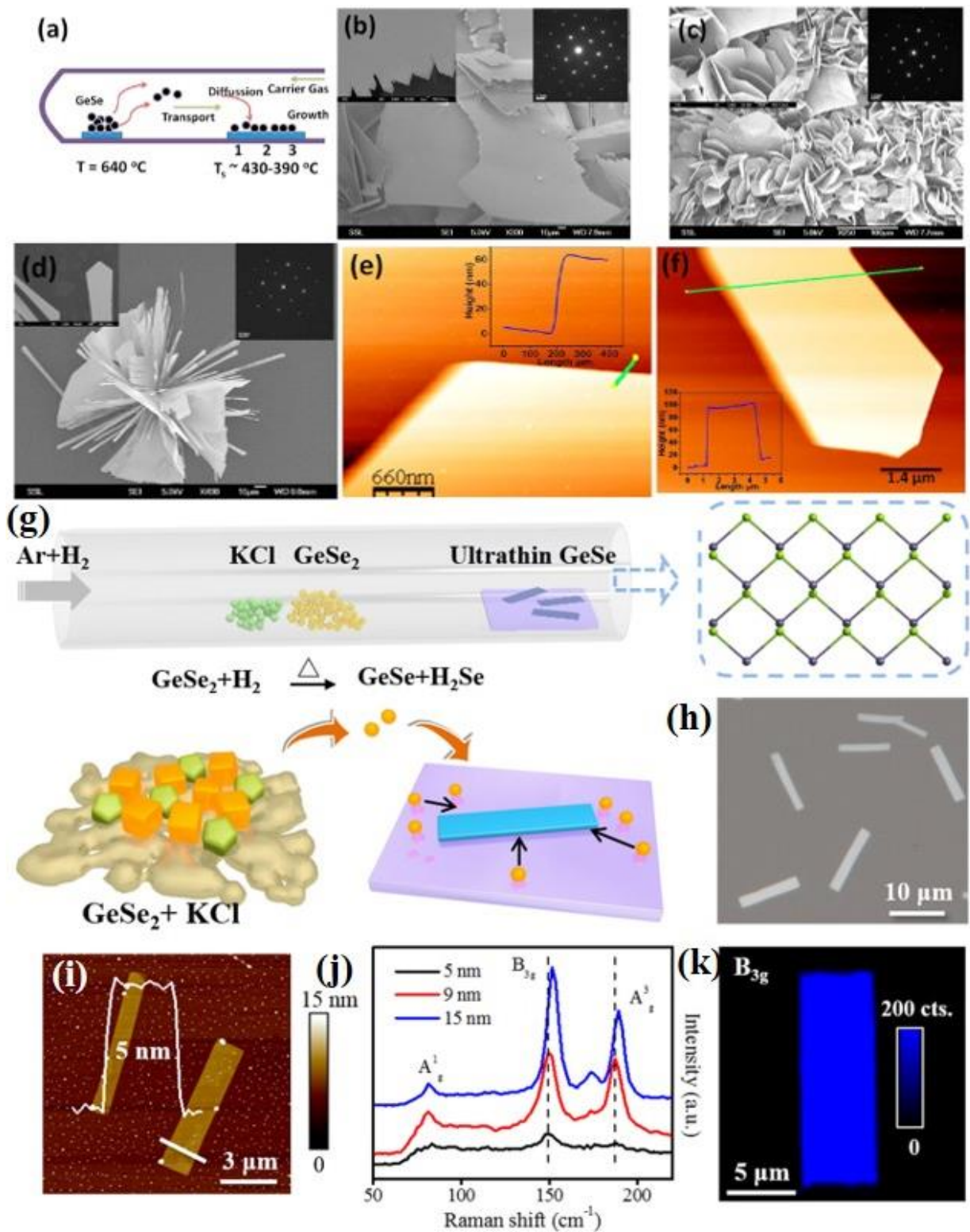

**Figure 10.** *a) Schematic CVD synthetic process. b-d) Scanning electron microscopic (SEM) images of CVD grown GeSe flakes on the areas indicated by 1, 2, and 3 in figure a, respectively (Insets: top right corners of b, c, and d are SAED patterns for representative flakes). The insets*

*panel at the left corners of b, c, and d are the magnified SEM images. e, f) AFM images of GeSe flakes and corresponding height profile (inset in e, f) of the flakes (the solid line in e, f). Reproduced with permission.[68] Copyright 2013, American chemical society. g) Schematic representation of the salt-assisted CVD with atomic structure of the layered GeSe. h) Optical image of as grown GeSe flakes. i) AFM image of GeSe flakes with height profile j) Raman spectra of GeSe flakes. k) Raman mapping ($B_{3g}$ mode) of GeSe flake. Reproduced with permission.[166] Copyright 2019, American chemical society.*

Moreover, the synthesized GeSe flakes are highly uniform and homogenous, which is reflected in respective Raman mapping of $B_{3g}$ mode (**Figure 10k**). In addition, a sequence of polarization-resolved Raman spectra for $A_g$ and $B_{3g}$ modes has been recorded, showing that the angular dependence on crystallographic directions of the NSs is identical with that of mechanically exfoliated GeSe flakes.[102]

PVD synthesis of orthorhombic SnS has been reported by Meng and co-workers.[70] The authors investigated the effects of growth temperature and pressure. During synthesis bulk SnS powder was evaporated into a horizontal single zone tube furnace, upon heating at 600-800°C for 10 minutes at a pressure around 20-300 Torr (**Figure 11a**). Mica sheets were used as substrates , which were placed downstream to the tube center at a distance of ∼8-20 cm from the evaporation area. A natural cooling process was employed to cool down the system. Throughout the experiment, Ar was used a as carrier gas with a constant flow rate. The synthesized SnS flakes on mica substrates exhibit quasi-rhombic shapes with a lateral size in the range from hundreds of nanometers to one micron (**Figure 11b, c**). However, the random nucleation of SnS resulted in a spatial non-uniformity of the grown flakes. (**Figure 11c, d**). The thinner NSs showed relatively rough surfaces and edges than the thicker ones, which was signified the incomplete crystal growth of SnS flakes. They further confirmed that the growth temperature and pressure significantly affect the domain size and grain boundary edge structure of synthesized 2D materials.[170, 171] High-resolution TEM images (**Figure 11e**) of the synthesized SnS showed a perfect rhombus lattice with fringe spacings of ∼0.29 nm, ∼0.4 nm and ∼0.43 nm, corresponding to the interplanar spacings of the (101), (002) and (200) planes of the orthorhombic SnS, respectively. The corner angle is measured to be 43°, which agreed well with the theoretical value. The vibrational properties of synthesized 2D SnS flakes with different thickness were monitored, showing that the SnS flakes exhibited four characteristic phonon modes, i.e. three $A_g$ and one $B_{3g}$, peaked at 95.5, 190.7, 216.8 cm$^{-1}$, and 162.5 cm$^{-1}$ respectively. It was identified that the Raman intensity and peaks deviate from the peak positions of $A_g$ and $B_{3g}$ upon lowering the vertical dimensionality. The strange spectral behavior in Raman spectrum (**Figure 11f**) of the thinnest NS (5.5 nm) was due to poor crystallinity (**Figure 11d**).

In another work, Tian et al.[172] have reported the synthesis of layered anisotropic 2D SnS via PVD. A lower sublimation temperature was adopted to evaporate the bulk SnS powder. In their typical synthesis, a two-zone tube furnace was used (**Figure 11g**). For the synthesis, the bulk SnS powder was heated at 530-560 °C, while a high purity Ar gas was used to carry the SnS vapor and deposit it onto mica substrates. (**Figure 11h**). The obtained NSs exhibited the orthorhombic crystal structure, lateral dimensions of 5-15 μm and minimum thickness of 6.3 nm (**Figure 11i, j**). TEM analysis showed perfect rhombus lattice fringes (**Figure 12k, l**) with d spacings of 2.90, 2.90, 4.32, and 4.01 nm, corresponding to [011], [011‾], [010], and [001] directions, respectively. Moreover, the measured corner angle (85°), between the [011] and [011‾] planes, well agreed with the theoretical predictions.[173, 174] While, the perfect rhombus lattice fringes in SAED pattern further confirmed the single crystallinity of the NSs. Finally the NSs exhibited the typical characteristic $A_g$ and $B_{3g}$ Raman modes, with the thicker NSs to show stronger and sharper Raman modes compared to the thinner counterparts. This confirmed the better crystalline quality of the thicker NSs, which was consistent with the edge morphology obtained from AFM images.

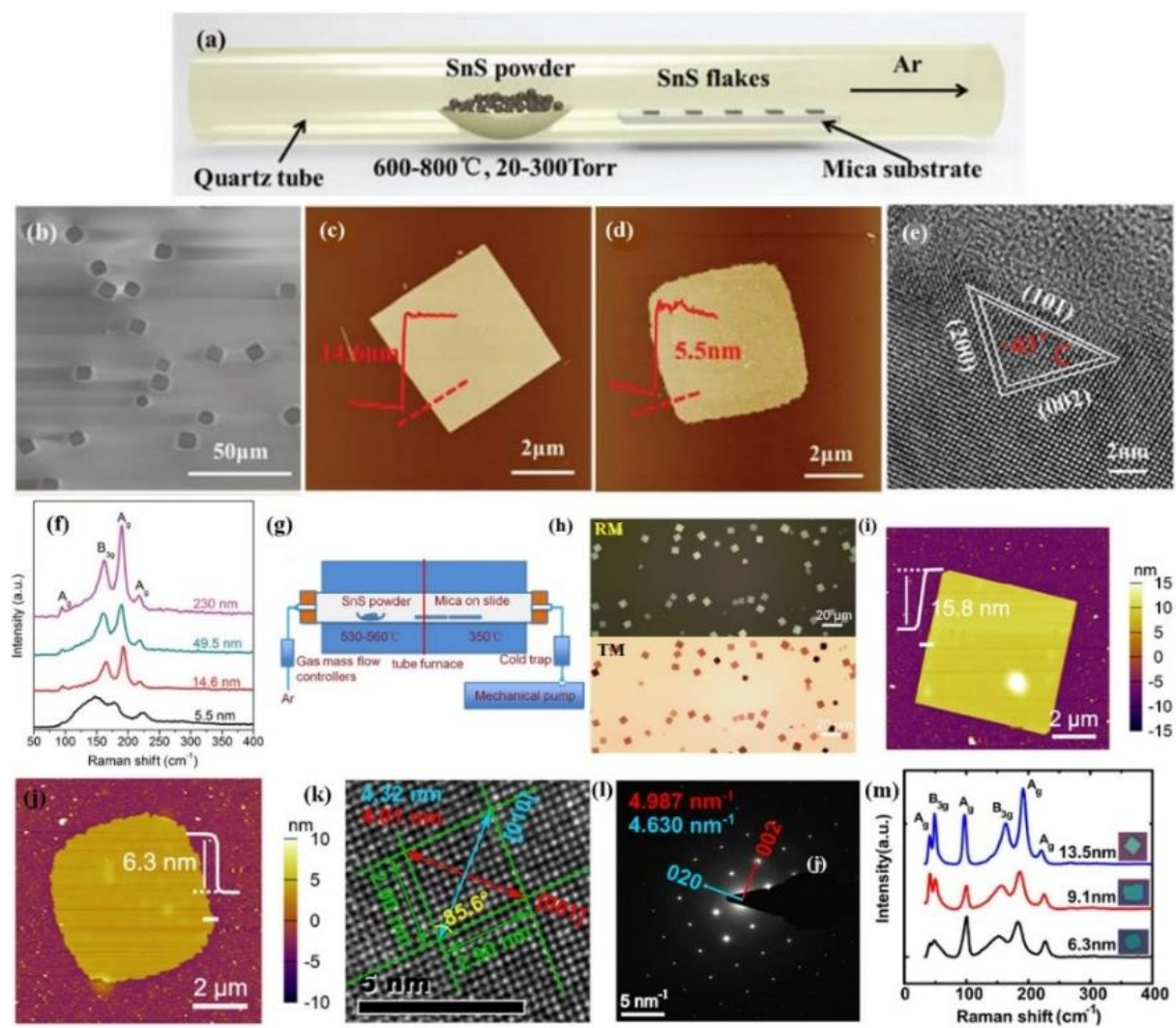

**Figure 11.** *a) Schematic of PVD growth for anisotropic SnS flakes; b) SEM image of the anisotropic SnS flakes synthesized at 600 °C. c, d) AFM images and height profile of anisotropic SnS flakes. e) HRTEM image of the SnS flake; f) Raman spectra of the SnS flakes with different thicknesses. Reproduced with permission.[70] Copyright 2016, Royal Society Chemistry. g) Schematic representation of the PVD growth system; Characterization of synthesized 2D SnS nanoplates (h) Optical microscopy images on mica substrates (RM and TM correspond to reflection transmission mode of microscopy). i, j) AFM images with height profile. k) High resolution TEM image. l) SAED pattern corresponding to the flake shown in k.*

*m) Raman spectra of SnS nanoplate with different thickness. The corresponding optical microscopic images are shown in the inset. Reproduced with permission.[172] Copyright 2017, American chemical society.*

The synthesis of few layer 2D SnSe via PVD has been reported by various groups.[61] Zhao et al.[163] have realized the synthesis of single-crystal SnSe NSs on mica substrates in a controlled manner. A 15.8 nm thick SnSe NSs having orthogonal lattice fringes with well define lattice spacings (0.30 nm), and intersect angle (92°) of the crystal planes were identified, which further confirmed the orthorhombic SnSe crystal structures.[126, 175] In PVD method, the temperature mainly controls the evaporation quantity of bulk SnSe, while the pressure affects the nucleation and growth process. With the evaporation temperature and pressure fixed at 500°C and 70 Torr, the deposition temperature was varied (from 340-390°C), and as a result controlled synthesis of SnSe NSs was achieved.[176] A few years ago, Xu et al.[164] have followed a similar synthesis process to that reported by Zhao et al.,[163] however the sublimation and the deposition temperatures were kept higher. In particular, the sublimation, deposition temperature of SnSe and argon flow rate was set at 650°C (in 17 min), 350-450 °C, and 200 standard cubic centimeters per minute (sccm). (**Figure 12a, b**). Following this process, SnSe NSs with a uniform surface were grown. However, a distribution of thickness (10 nm to 210 nm) was found, which can be controlled through the growth conditions. Interestingly, the rectangular NSs showed clear orthogonal lattice fringes with two similar lattice spacings of 0.30 nm. The measured angle (94°) between two crystallographic planes is well matched with the orthorhombic SnSe crystal structure.[175, 177] The observed orthogonal structure together with the orthogonal symmetry (**Figure 12c**) confirmed the single-crystal nature of the synthesized SnSe NSs. However, the corner angle was obtained to be 94°, which differ from that of other reports.[163] In fact, this discrepancy originates from the impact of different value of structural anisotropy (along armchair and zigzag direction) during the growth.

Owing to the phase transition occurred during CVD growth the substrate temperature critically affects the stoichiometry in SnSe (i.e the Sn:Se ratio),[178-180] Recently, Wang et al.[178] used a proper proportion of $SnO_2$ and Se precursors to synthesize 2D SnSe or $SnSe_2$ NSs with lateral sizes of few micrometers. Specifically, the Sn:Se stoichiometry was varied by changing the weight of Se from 50 to 500 mg. It was revealed that the shape of as grown sample switches from square to truncated triangle (**Figure 12 d, e**). As a consequence, the phase controlled growth of 2D tin selenides is possible by adjusting the nominal Sn:Se ratio. The AFM height analysis of SnSe NSs revealed a quite flat top surface and thickness of 59.8 and 95.1 nm (**Figure 12e**). On the other hand, the obtained thickness of the different-shape- $SnSe_2$ flakes grown were ~67.7 nm and ~20.8 nm, respectively. In order to shed light on the CVD process a simple growth mechanism was discussed.

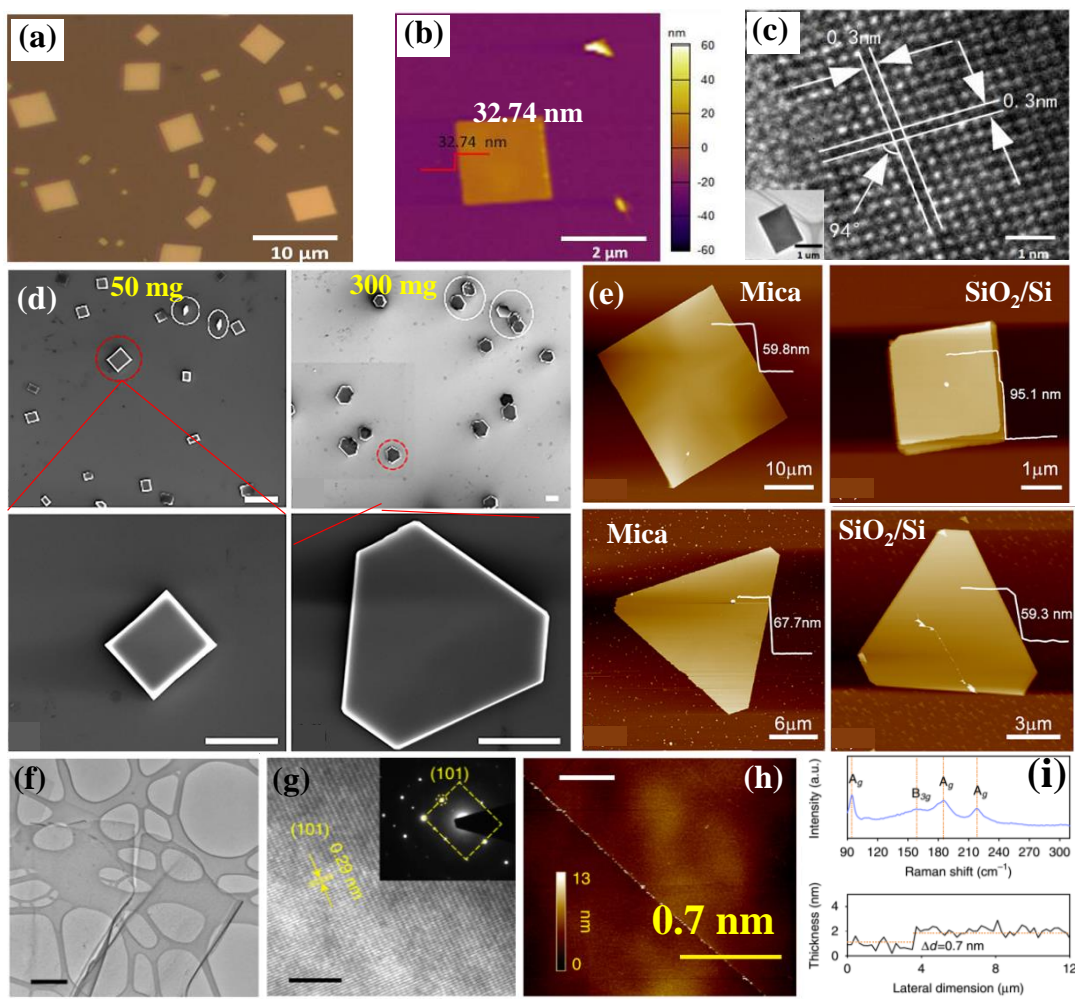

**Figure 12.** a) *Optical image of SnSe NS on mica substrates (lateral size distribution of 2 μm to 8 μm). b) AFM image with height profile (inset solid line, height 32.74 nm). c) HRTEM image of a NS (scale bar is 1 nm); a low-magnification TEM image is shown in the inset. Reproduced with permission.*[164] *Copyright 2017, American Chemical Society. d) SEM images of as grown flakes on SiO2/Si substrates with 50 mg and 300 mg Se powders (Scale bar is 20 μm (top panel) and 10 μm (bottom panel). e) AFM images of as grown and of different shape flakes on mica and SiO$_2$/Si. Reproduced with permission.*[178] *Copyright 2018, Elsevier. f) TEM image of SnS monolayer synthesized* using a liquid metal exfoliation method. *Scale bar is 500 nm. g) HRTEM fringe pattern. Inset: SAED pattern. Scale bar is 5 nm. h) AFM image of the SnS monolayer, and i) Raman spectrum and AFM height profile of the SnS monolayer. Scale bar 8 μm.*

In particular, when the bulk $SnO_2$ powders are heated to react with Se vapor, and the reaction forms SnSe or $SnSe_2$ via the following reaction paths:

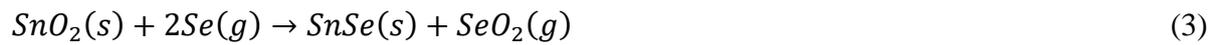

$$SnO_2(s) + 2Se(g) \rightarrow SnSe(s) + SeO_2(g) \tag{3}$$

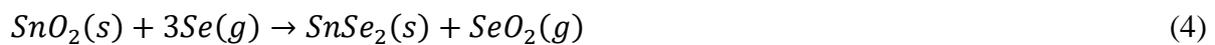

$$SnO_2(s) + 3Se(g) \rightarrow SnSe_2(s) + SeO_2(g) \tag{4}$$

When the amount of Se is low, the first chemical reaction takes place giving rise to the formation of SnSe. On the other hand, upon increasing the weight of Se precursor, the second chemical reaction is dominant and $SnSe_2$ NSs are formed.

Very recently, Khan et al.[181] have introduced a novel bottom up synthesis method to obtain SnS monolayers. In particular, a molten droplet of Sn was exposed to an anoxic atmosphere containing a sulfur source at 350 °C. As a result, the surface forms a sulfide skin in a self-limiting Cabrera-Mott reaction. Then liquid metals are employed to exfoliate ultrathin SnS sheets and to transfer them onto the desired substrates. Using this method, SnS nanosheets with large lateral dimensions can be attained, with a highly-crystalline orthorhombic structure (**Figure 12f,g**). The ultrathin SnS layer thickness obtained was measured to be 0.7 nm (**Figure 12h**), which was further confirmed by Raman spectroscopy (**Figure 12i**). The characteristic features are agreed well with the phonon mode in a thin layer SnS. This synthesis method may suitable for the isolation of the MMCs single layer.

Another bottom up approach is the wet chemical synthesis (WCS) method. This method comprises hot injection, one pot synthesis, hydrothermal intercalation and sequential deposition.[4, 61, 126, 182, 183] As a fast and low-cost method for producing nanomaterials for large scale industrial application, the WCS of MMCs nanosheets has drawn attention recently, to produce SnS, SnSe, and GeSe nanosheets[150, 184]. In a typical wet synthesis of colloidal MMC nanosheets using the hot injection method, the group-IV elements, such as $SnCl_2$, $SnCl_4$ and $Ge_4$, are used as an inorganic halide metal source.[150, 184] On the other hand, organic materials, such as dodecanethiol, trioctylphosphine selenide (TOP-Se) and thioacetamide are used as group-VI elements. Both reactants are mixed together with organic solvents and heated up to the reaction temperature. Subsequently, the obtained products are redispersed into an organic solvent and centrifuged to isolate the required nanosheets. In 2013, Li et al.[184] has adopted onepot synthesis as a WCS method to synthesize ~300 nm wide and 2ML (~1nm) thick SnSe nanosheets. In another report Schaak and co-workers[150] has adopted one pot synthesis method to prepare GeS and GeSe nanosheets. The obtained GeS nanosheets were 5 nm-thick with (2-4) µm times (0.5-1) µm average lateral dimensions. For GeSe a larger range of thicknesses was achieved (5- 100 nm). Apart from the aforementioned approached, much larger MMC flakes has been synthesized via WCS.[182, 183]

**Table 5.** *Comparison of the typical methods used for the synthesis of MMCs nanosheets*

| Method | Brief description of the synthesis method | Materials | Lateral dimension | Vertical dimension | Advantages | Limitations | References |
|---|---|---|---|---|---|---|---|
| **Mechanical exfoliation** | Scotch tape is used to peeled off thin layer from bulk crystal. Gold tape, polymer matrix is use to transfer thin layer on required substrates | GeS | Few tens of nm | 8 nm | Simplicity, high crystal quality, low defects | Low exfoliation yield, repeatability in size, reproducibility in layer number and the large area uniform flake | [107] |
| | | GeSe | Tens of µm | 33 nm | | | [101] |
| | | SnS | several µm | 4.3 nm | | | [57] |
| | | SnSe | Tens of µm | 90 nm | | | [105] |
| **Liquid phase exfoliation** | Bulk crystals or powders are dispersed in a suitable solvent and are subjected to ultrasonication for a certain amount of time. Suspension is centrifuges to isolate the ultrathin nanosheets (NSs) | GeS | - | 1.3±0.1 | Solution processed, large scale bulk production, high yield, low cost, simplicity | Thickness control, relatively smaller lateral dimension, proper choice of solvent | [143] |
| | | GeSe | 50-200 nm | 2 | | | [141] |
| | | SnS | 170 nm | 1.11 nm | | | [110] |
| | | SnSe | 150 nm | 2-10 nm | | | [149] |
| **Chemical vapor deposition** | One of the reliable method to produce 2D material for electronics. The CVD | GeS | 1.5-20 µm | 10 nm | Large scale lateral size, precise controllable | High temperature, ambient environment, high | [159] |

| | | | Lateral | | | | Ref |
|---|---|---|---|---|---|---|---|
| process employs reactive precursors and high vacuum, in which the precursors react and/or decompose on the surface of the substrate at high temperature to form ultrathin flakes | GeSe | Few µm | 5 nm | Thickness and lateral dimension, less defects | vacuum, Relatively complicated recopies, costly | [166] |
| | | SnS | Few µm | 5.5 nm | | | [70] |
| | | SnSe | 1-6 µm | 6- 40 nm | | | [163] |
| **Wet chemical synthesis** | Chemical method, surfactants or polymers assisted direct synthesis process | GeS | 2-4 µm | 5 nm | Solution processability, high production yield | Defects, surfactants on the surface | [150] |
| | | GeSe | - | 5-100 nm | | | [150] |
| | | SnS | 8 µm | 7 nm | | | *[182]* |
| | | SnSe | 300 nm | 1 nm | | | *[184]* |

**Table 5** summarizes the comparison of the various synthesis methods applied for the synthesis of MMC flakes. In general, there is currently no method that can produce high crystal quality and ultrathin (down to monolayer) coupled with high-lateral-dimension nanosheets. For example, the mechanical exfoliation and CVD/PVD techniques can produce as large as few µm to tens of µm in lateral dimensions MMCs, while it is difficult to isolate nanosheet thicknesses below 4 nm. On the other hand, the LPE and WCS methods can produce ultrathin, even monolayer flakes, however, the lateral dimension is very small and thus not favorable for optical and electronic properties investigations. Besides this, mechanically and LP exfoliated MMCs exhibit high crystal quality, contrary to the CVD/PVD grown ones.[172] This is primarily due to the strong interlayer coupling energy (armchair/zig-zag) in MMCs, which restricts the isolation of a single layer with high crystallinity. In this context, there are a few possible ways to overcome the different limitations. As far as the ME and PVD/CVD techniques, by proper choice of the deposition substrate and optimized protocols, one could make the lateral growth stronger than the vertical one, and as a result thinner and larger flakes can be attained. On the other hand, the post etching of mechanically exfoliated MMC flakes via chemical routes and/or laser processing can overcome the thickness limitation and give rise to large area monolayer flakes.

## 4. Applications: Theory and experiment prospect

Since the isolation of single layer graphene a new era of 2D electronics has begun. The semimetalic nature with gapless band structure of graphene imposes important limitations in electronic, optoelectronic and photonic applications. However, 2D MMCs exhibit a wide range of direct or indirect bandgaps. More important, MMCs are highly- anisotropic in nature, as they exhibit in-plane structural anisotropy along the armchair and zigzag crystalline directions. Such structural anisotropy is revealed in electronic, optoelectronic and photonic response. In the following, we will present an overview of the very recent advances in electronic, optoelectronic and photonic applications of MMCs, giving emphasis in valley polarization (VP) and second harmonic generation (SHG).

### 4.1 Electronic devices exhibiting anisotropic response

In-plane anisotropy found in a layered BP open up a new horizon in 2D MMCs research for emerging nanophotonic and optoelectronic device applications.[19, 185] Following such first studies, in-plane phosphorene-analogous group IV-VI MMCs have received significant interest due to their anisotropic optical and electronic response.[100, 103, 113, 117, 186] In wavy structured MXs, the effective mass, dielectric constant and refractive index are entirely different along armchair compared to zig-zag direction. Stereochemically active lone pair electrons in 5s (e.g. Sn) or 4s (e.g. Ge) and different electronegativity in chalcogens (e.g. S or Se) play a crucial role in such anisotropy. This interesting feature critically affects the exotic optical and electronic behavior of such materials, therefore adds a new dimension to their optoelectronic properties and stimulates the development of angle-resolved photonics and optoelectronics.[129]

Accordingly, understanding the effect of electrical and opto-electrical anisotropy in MMCs has evolving rapidly in the recent years.

Several groups had investigated the electrical anisotropy in few layer GeS NSs.[100, 187] Matsuda and co-workers,[187] in particular, introduced a highly polarization sensitive and broadband photodetector (**Figure 13a**), based on multilayer germanium sulfide (GeS). A 45 nm GeS flake was used to fabricated a Field Effect Transistor (FET) device. Photoresponse was controlled via tuning of the gate bias voltage ($V_g$) and the light intensity. The transfer characteristics ($V_g$-I) of FET device was exhibited a p-type behavior (**Figure 13b**), and the obtained carrier mobility was measured to be $1.6 \times 10^{-3}$ cm$^2$ V$^{-1}$ s$^{-1}$. A high on/off current ratio ($>10^4$) was also recorded at $\pm 40$V. Moreover, current hysteresis appeared in the voltage sweeping, in which trap/defect states play a dominant role. The anisotropic crystal structure of GeS, causing its strong linear dichroism,[107] was investigated through measuring of the photoresponse of the GeS photodetector with highly polarized light. It is found that the measured photocurrent (at $\lambda \sim 750$ nm) strongly depends on the polarization angle ($\theta$) of incident light (**Figure 13c**). The photocurrent sensitivity (polar plot of **Figure 13d**) reaches its maximum value at $0^o$ and its minimum at $90^o$ polarization, corresponding to armchair and zig-zag directions, respectively. This strong electrical anisotropy is perfectly consistent with the anisotropy in optical absorption. As a consequence, a GeS-based linear dichroic photodetector was demonstrated with a dichroic ratio of 1.45. In another study, Li et al.[100] presented a photodetector fabricated using a GeS NS with a thickness of 28.7 nm (**Figure 13e, f**). It is shown that the angle-resolved photocurrent (**Figure 13g**) presented in a polar plot changed dramatically with incident light.

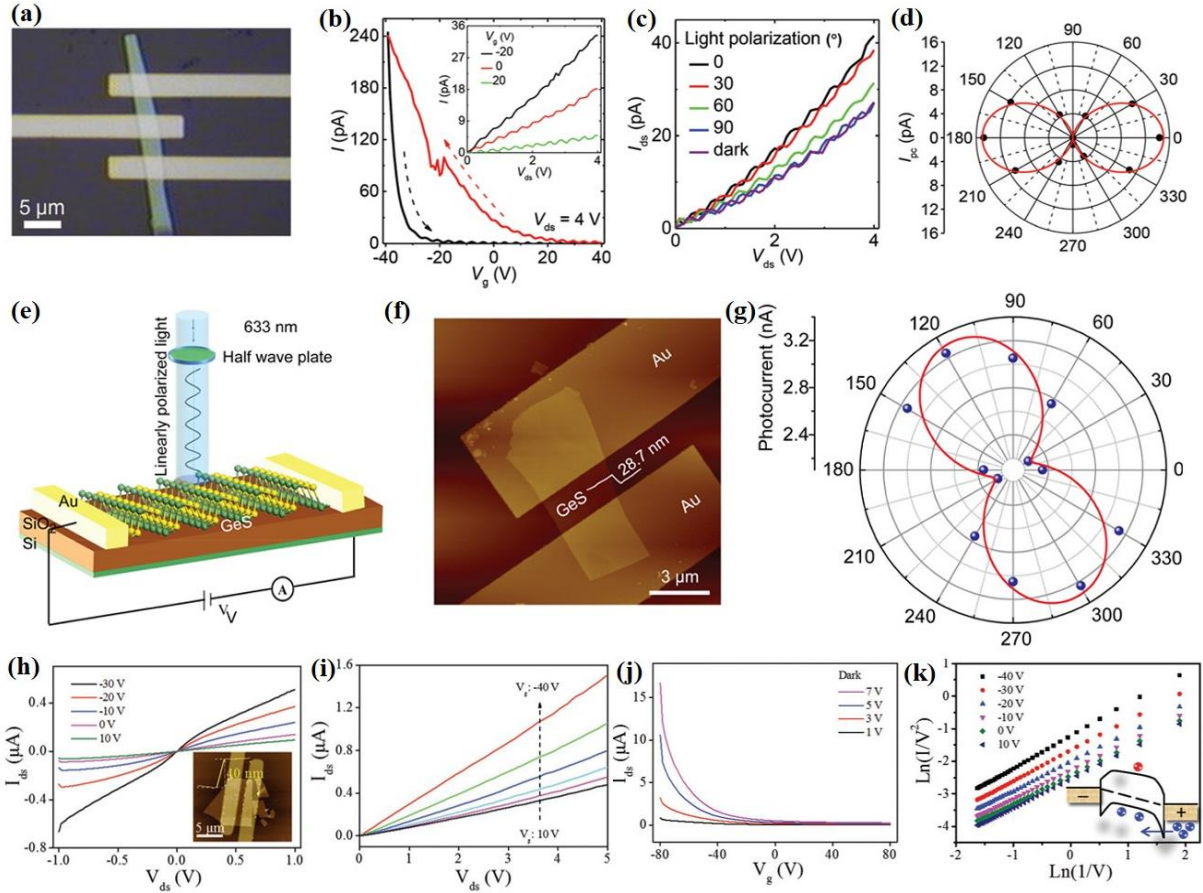

**Figure 13.** *a) Optical image of fabricated GeS photodetector (45 nm thick GeS); b) Gate bias voltage dependent transfer characteristics. Sweeping direction (black and red dotted arrow) indicating the forward and reverse voltage scans respectively (-40 to +40V and +40 to -40V). The inset shows $V_{ds}$-I characteristics at various back-gate voltages. c) Incident light (λ=750 nm) polarization dependent $I_{ds}$-$V_{ds}$ characteristics of a GeS FET device (P=32 μWcm⁻²). d) Polar plots of photocurrent. The angle θ is the polarization direction of the incident light relative to the armchair direction. The solid red curve represents the cos2θ function. a)-d) Reproduced with permission.[187] Copyright, 2017, Royal Society of Chemistry. e) Schematic diagram of a GeS phototransistor. f) AFM image of the device (scale bar: 3 μm). and g) Polar plot of photocurrent as a function of the polarization angle. e)-g) Reproduced with permission [100] Copyright 2019, American Chemical Society h) The $I_{ds}$-$V_{ds}$ characteristics of a GeSe transistor at different gate voltages. The AFM image of the device and height profile is shown in the inset (thickness and scale bar: 40 nm and 5 μm respectively). i) Device characteristic*

*curves at different gate bias voltages. j) Transfer characteristics of the device. k) Direct tunneling plots at different gate voltages. The inset shows an illustration of carrier tunneling. Reproduced with permission.[188] Copyright, 2018, Wiley-VCH.*

Hu and co-workers[113] first reported on the anisotropy in optical and electrical behavior of CVD grown few-layered GeSe. Following this study, Zhai and co-workers[188] reported on back-gated FETs from mechanically exfoliated GeSe NSs of 40 nm thickness. The characteristic I-V curves were linear (**Figure 13h, i**), indicating an Ohmic contact between the electrodes and the GeSe NSs. Besides this, the transfer characteristics exhibited a p-type semiconducting behavior (**Figure 13j, k**). A high on off/ratio ($\approx 10^3$) and hole ($0.8$ $cm^2V^{-1}s^{-1}$) mobility at room temperature was measured, which are comparable with other metal dichalcogenides such as $SnS_2$ and $SnSe_2$.[189-191] Furthermore, the photoresponsive behavior of the fabricated FET exhibited a deviation from the linear relationship (**Figure 14a, b**) under illumination (532 nm intensity of 0.42 mW $cm^{-2}$). It is shown that, a Schottky emission charge carrier transport mechanism takes place due to the large density of photoinduced charge carriers under illumination (**Figure 14c**). This phototransistor exhibited excellent stability in both photoresponse and decay rate. An anisotropic optoelectronic behavior was also investigated on such phototransistors (**Figure 14d-f**). In particular, the normalized photocurrent was demonstrated three periodic peaks at 0°, 180° and 360° with two valleys at 0° and 270°, while a strong polarization sensitive photodetection with peak to valley ratio ~1.3 was obtained. A strong electrical anisotropy was also measured in a multi-terminal FET device in which the electrodes are placed at an angle of 45°. Moreover, significant impact of anisotropy in angle-resolved hole mobilities at different temperatures was recorded (**Figure 14f**). Indeed, the calculated hole mobility along armchair and zig-zag direction is $\approx 6.03$ $cm^2V^{-1}s^{-1}$) and $\approx 3.25$ $cm^2V^{-1}s^{-1}$. It is observed that the effective mass, m, played a crucial role to the mobility ($\mu$), as $\mu \alpha \frac{1}{m}$. Based on theoretical calculations, the respective masses are $m_{armchair} = 0.16m_0$ and $m_{zigzag}$

$= 0.33m_0$, where, $m_0$ is the free-electron mass.[172, 192] Notablly, the theoretically calculated

mobility ratio ($\mu_{armchair}/\mu_{zigzag}$) is well matched with the experimental value. This ratio was

shown to increase from 1.85 to 3.15 upon decreasing the temperature from 300K to 60 K

(**Figure 14g**). Such anisotropy in $\mu$ along armchair and zigzag direction pave a new root to employ GeSe for novel optoelectronics applications. In addition, Liu et al.[193] have also investigated similar anisotropic electronic properties in GeSe devices. **Figure 14h** presents the schematic diagram of the angle-resolved transport setup they used for electrical anisotropy measurements. The typical I-V characteristics exhibited a linear relationship in dark condition,

while the current significantly increased under laser illumination. A significant

photoresponsivity (7.05 A W$^{-1}$) and specific detectivity (3.04×10$^8$ Jones) was recorded, which

is highest value among recent reports. Furthermore, high photocurrent sensitivity to

polarization angle was observed (**Figure 14i**), namely a significant increase in photocurrent

(55%) upon using a 90$^o$ polarization angle (**Figure 14j**). Moreover, the polarized photocurrent

showed a correlation with the phonon scattering in zig-zag direction, namely when the incident

light is parallel to the zig-zag direction the LO phonons get excited, while the TO phonon get

excited when it is parallel. The LO phonons, in particular, exhibited relatively large scattering

cross sections to the charge carriers. The potential anisotropy in the electrical behavior of SnS

NSs was investigated by Xue and co-workers.[172] For this purpose, PVD-grown SnS

nanoplatelets was used to fabricate FET devices. The anisotropy in electrical transport was

investigated on devices with cross-Hall-bar structure (**Figure 15a**). The temperature dependent

transfer characteristics of the fabricated FET devices along zig-zag and armchair directions

were subsequently recorded. The hole mobility of the devices was calculated using the

equation:

$$\mu = \frac{L}{W} \frac{dG}{C_g dV_g} = \frac{d\sigma}{C_g dV_g} \tag{5}$$

where, L and W are the length and width of the channel, G is the conductance, $C_g$ is the gate capacitance per unit area, and $\sigma$ is the conductivity along armchair or zig-zag direction (**Figure 15b**).

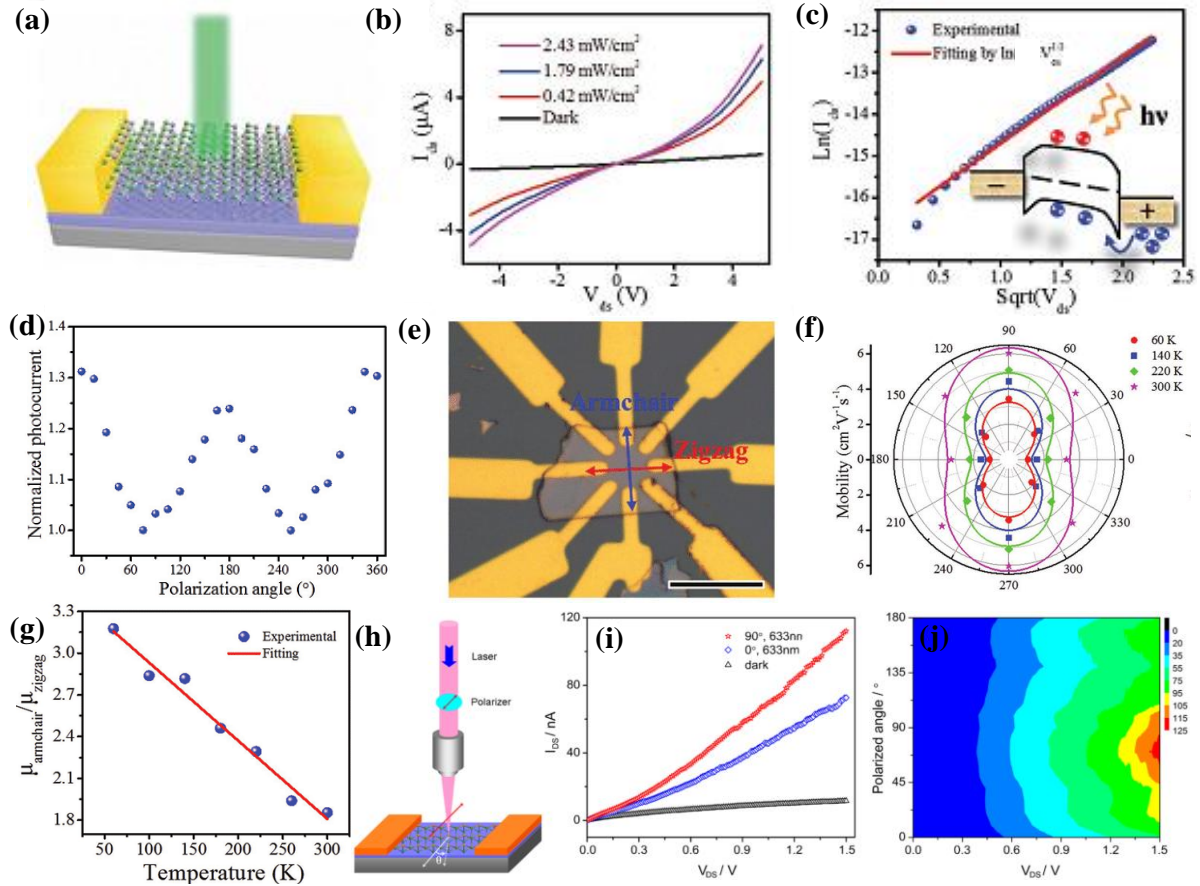

**Figure 14.** *a) schematic of a GeSe photodetector. b) I-V characteristicsc under dark and light illumination with different power intensities. c) Charge carrier transport via thermionic emission; Inset: Illustration of carrier transport under illumination. d) Normalized photocurrent as a function of the incident light polarization angle θ. e) Optical image of the device used for anisotropic electrical measurements (the scale bar is 10 μm). f) angle resolved mobility at different temperatures from 60 to 300 K. g) Temperature-dependent mobility ratio (μ<sub>armchair</sub>/μ<sub>zigzag</sub>) along the armchair and zigzag directions. Adapted from[188] Copyright 2018 Wiley-VCH. h) Schematic diagram of a setup used for angle-resolved transport measurements.*

It is found that the electrical mobility and conductivity values were higher in zig-zag direction, explained with the difference in the effective masses along the two principle axes. In particular, Vidal et al.[58] and Guo et al.[174] calculated that the effective masses along the zig-zag and armchair directions are very different, namely $m_{zigzag} = 0.21m_0$ and $m_{armchair} = 0.36m_0$. The estimated hole mobility ratio is $\mu_{zigzag}/\mu_{armchair} \approx 1.7$, which showed an excellent agreement with the experimentally-obtained value (**Figure 15c, d**). Furthermore, the activation energy was investigated along both directions in the crystal and found to be 46.5±1.7 meV and 43.1±0.8 meV along armchair and zig-zag direction, respectively. Very recently, Loh and co-workers[46] have reported gate tunable in-plane ferroelectricity in few-layer SnS FETs. A MBE grown few layer NS of 15 nm thickness was used to fabricate a large area SnS FET device exhibited cyclic I-V characteristics with -50 V back gate voltage and ±5V applied bias (**Figure 15e**). This characteristic was explained by the presence of polarized domains in SnS. Initially, with the positive bias voltage increasing from 0V, the device shows a low resistive state (LRS) with negatively-polarized domains. When the positive bias voltage exceeds +4.3V, the negatively polarized domains start to reverse to positively polarized ones and enter into the high resistive state (HRS)(from +4.3V to +5V). The HRS stays as the voltage is decreased to 0V. The corresponding coercive field, i.e the electric field when current peaks was ~10.7 kV/cm (at ±4.3V), which is remarkably smaller than the theoretically calculated value.[39] Such ferroelectric switching in multidomain SnS films was governed by the domain wall motion. In addition, an electrostatically tuned ferroelectricity has been appeared (**Figure 15f**) in SnS due to the carrier depletion. Such carrier depletion was tuned upon variation of the gate voltage.

The gate voltage-dependent polarization was recorded by applying a constant voltage sweep (1V) between the source and the drain (**Figure 15g**). The total polarization in the FET device can be calculated as $P(V) = \frac{1}{d}\int I(V)dV$, where, d is the channel length.

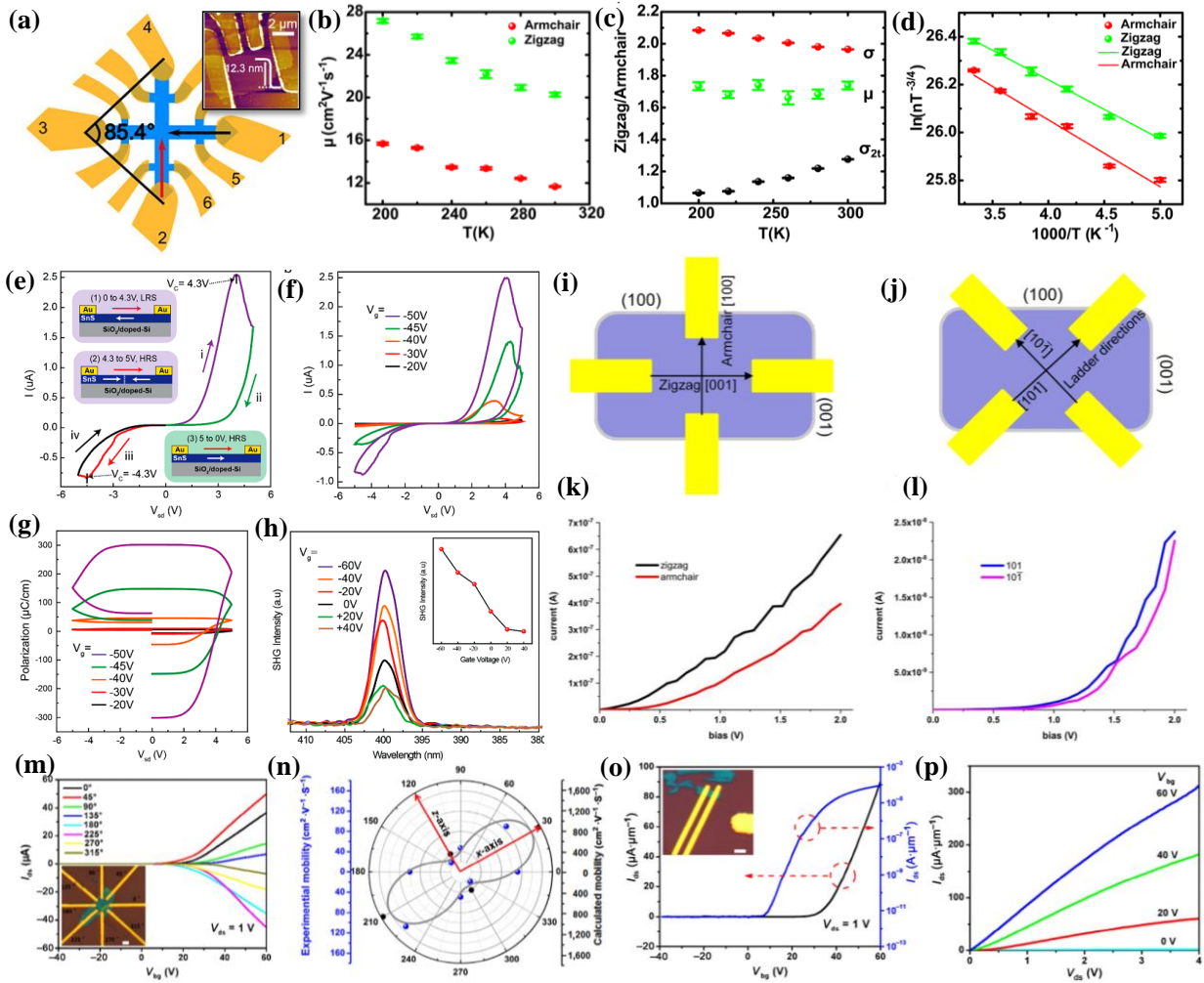

**Figure 15.** *a) Schematic of SnS (12.8 nm) FET device. The black and red arrows represent the anisotropic armchair and zigzag directions, respectively (Figure a). Inset: AFM image of the device. b) Mobility (μ) versus temperature at zero gate voltage. c) Temperature dependent ratio of μ, σ and two-terminal conductivity (σ₂ₜ). d) Carrier density as a function of temperature. Adapted from[172] Copyright 2017, American Chemical Society. I-V characteristics and SHG response of a FET based on few-layer SnS. e) The I-V hysteresis curve of the SnS FET device. f) I-V hysteresis curves of a lateral SnS memory device measured at different gate bias voltages.*

*g) Gate voltage vs polarization hysteresis curves of the memory device h) Second harmonic generation (SHG) peaks generated by exciting the back-gated SnS device with laser pulses. Inset: SHG peak intensity as a function of the gate bias voltage. Reproduced with permission[46] Copyright 2019, American chemical society. Schematic representation of the SnS NSs contacted along i) anisotropic directions j) isotropic directions. k) The current-voltage (I-V) characteristics along the armchair and zigzag directions l) The I-V characteristics along isotropic directions. Reproduced with permission[182] Copyright 2019, American chemical society. m) Room temperature transfer characteristics of SnSe FET based on few layered NSs. Inset is the optical image of the fabricated device. n) Normalized field-effect mobility of the SnSe FET shown in (m). Blue and black dots are experimental and theoretical data points respectively. The grey line is the fitting curve. o) Room-temperature transfer characteristics ($I_{ds}$-$V_{gs}$) of the transistors along the x-direction (at $V_{ds}$=1V). Inset is an optical image of the fabricated device p) Output characteristics with various back-gate voltages $V_{bg}$. Adapted from.[36]*

Such remnant polarization increases upon increasing the negative gate bias voltage, namely increasing the p-doping level in SnS. Gate bias voltage dependent second harmonic generation (SHG) was additionally observed in the SnS FET device. The SHG peak appeared at 400 nm with an excitation of 800 nm (**Figure 15h**). The corresponding peak intensity increased with applying more negative gate voltage. In another study, Klinke and co-workers[182] has demonstrated anisotropy in the electronic behavior of LPE-synthesized large area single-crystalline multilayered SnS NSs. Four contacts were used to measure the anisotropy in conductivity of the NSs along armchair and zig-zag crystallographic directions (**Figure 15i-k**). The obtained conductivity in armchair (001) and zig-zag (100) directions was measured to be 39 S/m and 65 S/m, respectively. The ratio of the conductivity along armchair and zigzag ($\sigma_{armchair}/\sigma_{zigzag}$) was equal to ~1.7, which is in very good agreement with reported values.[172, 183, 188] Besides this, the conductivity along the (101) and the (10Î) (ladder) crystal directions was also measured and found to be isotropic, namely 2.2 and 2.3 S/m respectively. (**Figure 15l**). These primary results showed the great potential of SnS as a building block for future electrooptic device application.

A much larger anisotropy in electrical performance was found in few layer SnSe NSs of 10 nm thickness.[36] The typical angle resolved electrical measurements have been carried out in a FET configuration with eight electrode devices spaced at 45°(**Figure 15m**). The transfer characteristics of the devices were collected with four pairs of diametrically-opposite bars as source/drain contacts, by applying an electrical field using a back-gate voltage $V_{bg}$ in the range of −40 to +60 V. The corresponding n-type field effect mobility was obtained by $\mu_x = [dI_{ds}/dV_{bg}] \times [L/(WC_iV_{ds})]$ and found to be highly angle-dependent (**Figure 15n**); the highest mobility revealed in the 30°/210° direction, while the lowest in 120°/300°. This behavior was closely fitted with the function $\mu_\theta = \mu_x\,cos^2(\theta - \phi) + \mu_z\,sin^2(\theta - \phi)$, where, $\mu_x$ and $\mu_z$ are the mobilities along the armchair (x) and zig-zag direction (z) respectively. The calculated anisotropy in electrical field effect mobility along x and z direction ($\mu_{x(max)}/\mu_{z(max)}$) was ~5.8, which is highest value than other anisotropic materials, including SnS (1.7),[172] ReS$_2$ (3.1),[194] and BP (1.5).[20] *ab initio* theoretical calculations offered a qualitative explanation for the angular dependence of the mobility, which was compared with experimental results. However, the theoretically predicted anisotropy in electrical mobility ratio is slightly lower (~4.3) than the experimental value. Besides this, the electrical performance was probed in two terminal SnSe devices (**Figure 15 o, p**); the highest FET mobility obtained was 254 cm$^2$·V$^{-1}$·s$^{-1}$, and an ON/OFF ratio exceeding ~2 × 10$^7$, which is higher value than that observed in the anisotropic BP.[195]

## 4.2 Energy conversion

In a typical solar energy conversion device, the donor materials dominate the main physical processes, including light absorption and exciton transport. In such devices 2D materials are widely used as in the active (donor) or in various interface/interconnecting layers for exciton generation or charge extraction.[196-204] A promising novel donor material should have direct band gap (1.2-1.6 eV), high absorption coefficient, high carrier mobility and low exciton binding energy, $E_B$. A low $E_B$ is favorable for the efficient separation of photo-generated charge carriers. Besides this, in a solar energy device, the power conversion efficiency (PCE) largely depends on the band alignment between donor and acceptor materials. Another promising alternative property is the bulk photovoltaic effect, attributed to a nonlinear optical response that yields net photocurrent in materials with net polarization.[38, 205] Owing to their outstanding light absorption properties, 2D TMDs (eg. $MoS_2$, $MoSe_2$, $WS_2$ and $WSe_2$) have been widely investigated for their application in solar energy conversion.[206] In a similar manner, in-plane anisotropic layered 2D MMCs have recently been investigated as potentially promising materials in photovoltaics.[38, 41, 42, 109, 172, 207-212]

Dai and co-workers[41] have designed and investigated a new photovoltaic system based on novel 2D MMCs. In particular, the photoresponse and the photovoltaic performance of GeS, GeSe, SnS and SnSe monolayers were evaluated by means of theoretical quantum transport simulations. The corresponding $E_B$, values have been additionally estimated to be 0.68, 0.52, 0.25 and 0.21 eV for GeS, GeSe, SnS and SnSe, respectively. These binding energies are much weaker compared with the respective ones calculated for SiC, GaN and $MoS_2$ monolayers (**Figure 16a**).[213] Notably, some Se based MMCs exhibit two times lower $E_B$ than TMDs, leading to easier separation of the photo-excited monolayer MMCs were calculated (**Table 6**)[31] and found to be comparable with 3D Si ($\varepsilon$ =11.9), 2D $MoS_2$ ($\varepsilon$ =7) and anisotropic BP.[214-216] The higher dielectric constant of selenium based MMCS comply with lower $E_B$ values. The photovoltaic performance of monolayer MMCs was evaluted with a two-probe device configuration (**Figure 16b**) and a large photocurrent was measured under illumination with 0.1 W cm$^2$ light. Moreover the obtained photocurrent along armchair direction was higher compared to the zig-zag one. Notably, in seleneum based monolayers, the induced photocurrents, namely 12.0 and 14.0 mA mm$^{-2}$ for Ge and Sn ones respectively, were higher than the sulfur based ones. In addition, a red shift in photocurrent was appeared in Selenides compared to Sulfides (**Figure 16c-j**). This important phenomenon is attributed to their lower band gaps.[43, 217] The photovoltaic performance was evaluated through a photoresponce coefficient ($R_{ph}$) and EQE. It is important that the $R_{ph}$ vaules in MMCs are much higher than those of MoS$_2$ phototransistors and three orders of magnitude higher than that of graphene detectors. As shown in **Table 6**, the corresponding EQE values are higher in Selenides than Sulfides, indicating that the former are more suitable as photovoltaic materials than the latter ones. Very recently Rengel et al.[38] have reported a large photocurrent shift ($\sim$100 $\mu$A/V$^2$) in single layer Ge and Sn MMCs. Such shift was correlated with the large spontaneous effective three-dimensional electric polarization of $\sim$1.9 C/m$^2$ in Ge and Se MMCs.

**Table 6.** *Summary of dielectric constant of different materials with structural in-plane anisotropy obtained from theoretical calculations.*[31, 214, 216, 218]

| Materials | Layer number/thickness | Dielectric constant ($\varepsilon$) | | EQE (%) | References |
|---|---|---|---|---|---|
| Si | 2 nm thick | 11.9 | | - | [215, 216] |
| MoS$_2$ | Monolayer | 7 | | - | [214, 216] |
| PB | | 7.47 (armchair) | 3.06 (zigzag) | - | [218] |

| | | | | |
|---|---|---|---|---|
| GeS | 8.7 (armchair) | 8.6 (zigzag) | 10.27 | [31, 41] |
| GeSe | 13.8 (armchair) | 14.7 (zigzag) | 25.43 | [31, 41] |
| SnS | 9.9 (armchair) | 10.0 (zigzag) | 22.01 | [31, 41] |
| SnSe | 12.5 (armchair) | 12.8 (zigzag) | 30.32 | [31, 41] |

In addition, Lv et al.[219] have systematically explored the electrical properties of bilayer GeSe with different metal electrodes. In particular, the interface geometry, electronic properties, band alignment, Schottky and tunneling barriers were calculated with Au, Ag, Al, Cu, Pt, and Ni, electrodes respectively. Among such metals tested, Au, Pt and Ni showed lower Schottky and tunneling barriers. In addition, a heterostructure of monolayer SnS with a bilayer GeSe has been constructed to investigate the resulting photovoltaic performance (**Figure 17a**). In which, the CBM and VBM energy of donor is much lower than the energy of acceptor and forming a type-II band alignment. This type-II band alignment is favorable for effective separation of photo-generated charge carriers. In such device the PCE was calculated to be 18% (**Figure 17b**), which is larger than the best certified efficiency of organic photovoltaic cells[220] and comparable with other heterostructures-based[221-223] solar cells.

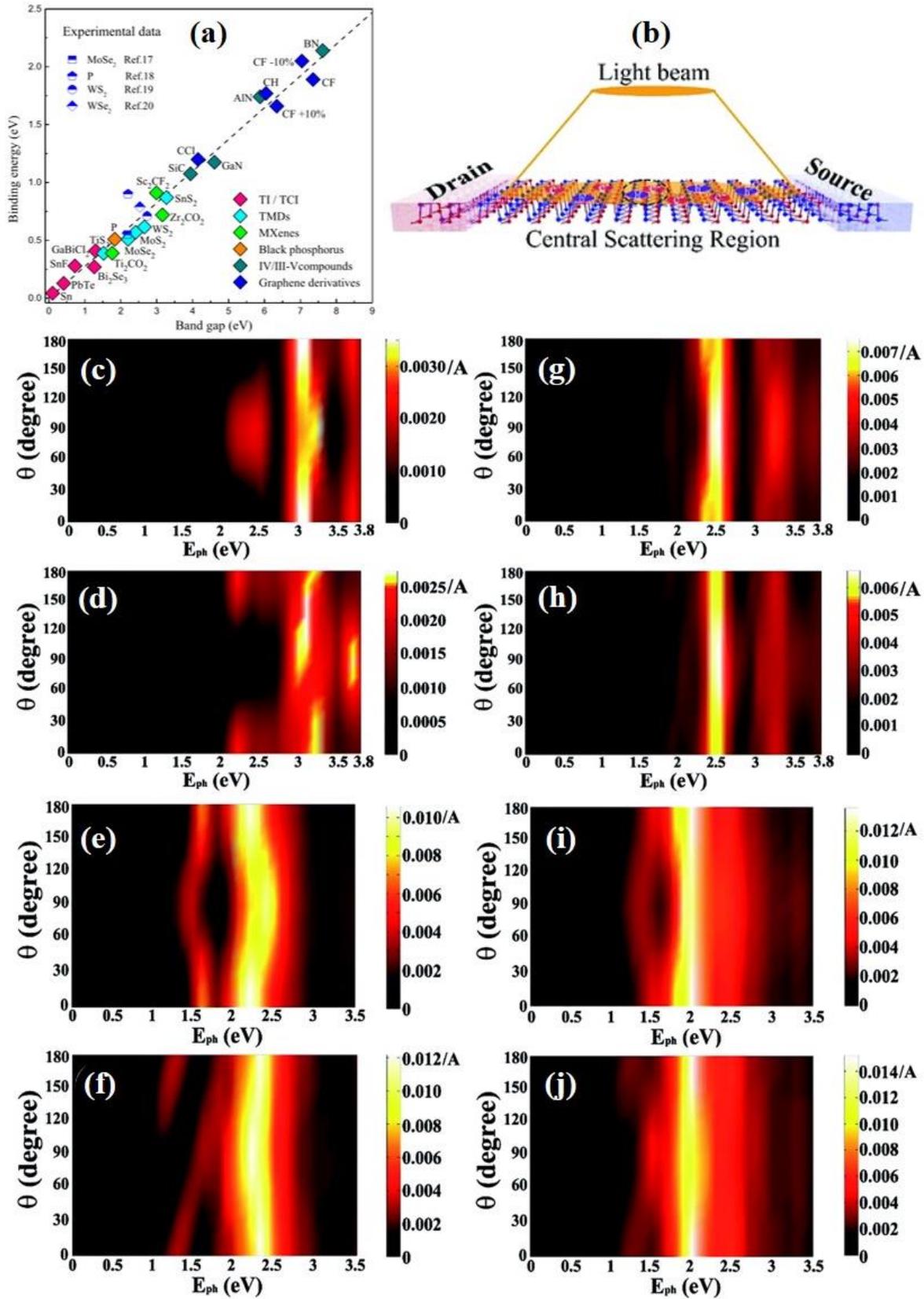

**Figure 16.** *a) Band gap vs exciton binding energy of existing 2D and bulk materials,[213] b) Schematic of two-probe monolayer devices. Photocurrent along armchair and zig-zag*

The extended absorption of MMCs in the IR range was exploited by Hou et al.[224] The authors fabricated bi-component heterojunction solar cells (**Figure 17c**) comprising an organic perovskite layer matched with a GeS, SnS, GeSe, or SnSe layer. The calculated absorption efficiency of the bi-component devices complied well with that of the combined absorption of the two components (**Figure 17d-g**). Notably, the device comprising a GeSe layer coupled with a $CH_3NH_3PbI_3$ one to form its active area, exhibited a strong absorption in the range of 300-1200 nm. Likewise, the corresponding external and internal quantum efficiency of the bi-component heterojunction solar cells were extended in the range of 300-1200 nm, contrary to those of the single perovskite layer cell, which were only limited in the range of 300-800 nm. Furthermore, the current density- voltage characteristic curves of the solar cells were presented (**Figure 17h**). It is shown that all the electrical parameters, including the short circuit current, $J_{sc}$ the open circuit voltage, $V_{oc}$ the fill factor (FF) and the PCE of the bi-component cell were much higher than those of the single-layer ones. In particular, the obtained $J_{sc}$, $V_{oc}$, FF and PCE values were 37.62 mA/cm$^2$, 0.76 V and 83.14%, 23.77% for the bi-component solar cell and 18.53 mA/cm$^2$, 1.02 V, 88.15% and 16.66% for the single perovskite cell, respectively. The significant rise of $J_{sc}$ observe in the bi-component cell can be attributed to the enhanced absorption, at higher wavelengths, by the GeSe layer. In another work Michael et al. [42] have investigated the performance of 2D piezo phototronic based solar cells. In such devices an external strain was applied to monitor the piezo phototronic effect. The strain-induced PCE of a monolayer SnS, SnSe, GeS and GeSe, can improve up to 18.2%, 16.6%, 7.0% and 15.4%, respectively.

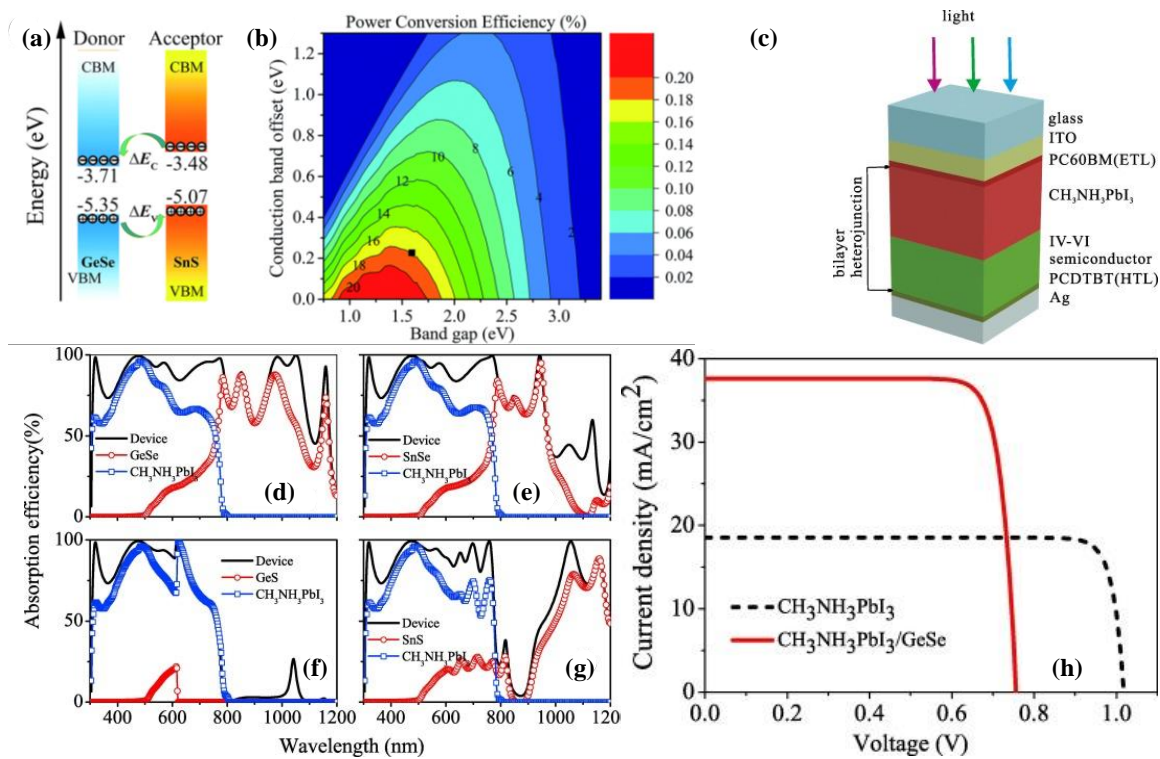

**Figure 17.** *a) The energy band alignments of 2D GeSe and SnS monolayers. b) Contour plot of the computed power conversion efficiency. Reproduced with permission.[219] Copyright 2018, The Royal Society of Chemistry. c) Schematic structure of a bi-component heterojunction solar cell comprising a perovskite (CH₃NH₃PbI₃) layer coupled with an MMC one to form its active area; Calculated absorption efficiency d) CH₃NH₃PbI₃/GeSe, e) CH₃NH₃PbI₃/SnSe, f) CH₃NH₃PbI₃/GeS, g) CH₃NH₃PbI₃/SnS. h) J-V curves of bi-component and single layer (perovskite only) solar cell. Reproduced with permission.[224] Copyright 2018, Elsevier.*

Another important application of MMCs is their catalytic activity.[225-228] In particular, the photocatalytic water splitting for energy conversion and the electrode catalyst (cathode) activity for energy storage have been theoretically investigated. In photocatalytic water splitting and electrode catalytic activity, both the electronic structure and the optical properties of the semiconductor play an important role. When an incident light falls on a semiconducting catalysis material, an electron gets excited to CB and generate electron-hole (e-h) pairs. Subsequently the electrons then take part in hydrogen evolution reaction (HER), while the holes cause an oxygen evolution reaction (OER), which are expressed as, $H_2:2H^+ + 2e^- \rightarrow H_2$ (HER) and $O_2:H_2O + 2h^+ \rightarrow 1/2O_2 + 2H^+$ (OER), respectively. The minimum $E_g$ required for these two processes to simultaneously take place is 1.23 eV.[229] More interestingly, Zhang and co-workers[230] have reported that an indirect band gap material is more suitable for photocatalytic activity. Most of MMCs exhibit larger band gaps (direct or indirect) than the minimum energy required for the water splitting reaction (>1.23 eV).[226] In particular the, GeSe, SnS and SnSe exhibited larger over-potential, which was behaving as a potential material for HER. On the contrary, SiS, SiSe, and GeS exhibit good photocatalytic activity in basic and acidic conditions. Apart from the photocatalytic activity, MMCs are predicted to be promising as cathode catalyst materials in $Li-O_2$ batteries.[227] However, the impact of in-plane anisotropy of MMCs in catalytic activity has not yet been investigated.

## 4.3 Valleytronics

The term "valleytronics" comes as the combination of the words 'valley' and 'electronics' and describes the ability to tune the carrier conduction in the valleys of a material's band structure. The idea was initially applied for the development of new quantum computing devices,[231-234] which use the valley polarization as a means to store and/or carry the quantum information.[231, 235] Neto and co-workers[37] theoretically predicted the valley properties and the optical selection rules in MMCs. In a particular, the 2D form of SnS was studied via a combination of *ab initio* calculations and 'k·p; theory. In monolayer SnS two pairs of valleys were identified, placed along two perpendicular axes, which can be selected exclusively with linearly polarized light and can be separated using nonlocal electrical measurements. Thereafter, Park and co-workers[39] investigated the multi-stability of the direction of the puckering of monolayer MMCs using first principle calculations. It is found that the monolayer of SnS and GeSe exhibit two inequivalent valleys in momentum space, which are dictated by the puckering orientation along armchair or zigzag direction. These valleys can be excited selectively using linearly polarized light.

The valleytronic behavior of low-dimensional materials can be probed through the variation of different experimental conditions, such as cryogenic temperature and/or strong electric and magnetic field. However, such requirements inevitably pose a plethora of practical challenges that create a high barrier in advancing the technology towards practical applications. Very recently, the realization of valley polarization at room temperature in MMCs was revealed. In particular, Lin et al.[51] demonstrated the direct access and identification of different sets of valleys in bulk in-plane anisotropic SnS (**Figure 18a**). The valley selectivity was recorded via the photoluminescence (PL) setup for SnS (**Figure 18, top panel**) and the corresponding polarization degree of the valleys was determined by[236]

$$P_{2D} = \frac{I(\sigma_-) - I(\sigma_+)}{I(\sigma_-) + I(\sigma_+)} \tag{6}$$

where, $I(\sigma_{+/-})$ represents the left (+) and right(-) circularly polarized PL intensity. Two peaks of SnS PL spectra were recorded at 817 nm (A) and 995 nm (B) (**Figure 18b**) respectively. To account for the valley polarization, the PL spectra were recorded upon rotating the sample with respect to the incident and detection polarizations (**Figure 18c, d**), that were parallel each other. The polarization degree called intervalley polarization,

$$P_{\text{intervalley, } \Gamma X} = \left( \frac{I_{\Gamma X \parallel}(\theta = 90^0) - I_{\Gamma Y \parallel}(\theta = 90^0)}{I_{\Gamma X \parallel}(\theta = 90^0) + I_{\Gamma X \perp}(\theta = 90^0)} \right) \tag{7}$$

$$\text{and } P_{\text{intervalley, } \Gamma Y} = \left( \frac{I_{\Gamma Y \parallel}(\theta = 0^0) - I_{\Gamma Y \parallel}(\theta = 0^0)}{I_{\Gamma Y \parallel}(\theta = 0^0) + I_{\Gamma X \parallel}(\theta = 0^0)} \right) \tag{8}$$

where, $I_{i \parallel}(\theta)$ stand for the PL intensity under parallel polarization for the $i$ valley. It is observed, that the (A)-peak maximizes at the polarization that minimizes the (B) peak, and vice versa. Such 90°-phase shift between the peaks signifies the selectivity of two valleys. The corresponding intervalley polarization degrees of the 817 (lying on ΓY axis) and 995 nm (lying on ΓX axis) peaks, calculated using the formula 7 & 8 were 92% and 62%, respectively. Unlike other 2D materials, SnS has a unique valleytronic system, which relies on the non-degeneracy of the valleys. In such system the valley polarization degree, between valleys and within a valley, can be assessed by,

$$P_{intravalley, \ \Gamma X} = \left( \frac{I_{\Gamma X ||}(\theta=90^0) - I_{\Gamma X \perp}(\theta=90^0)}{I_{\Gamma X ||}(\theta=90^0) + I_{\Gamma X \perp}(\theta=90^0)} \right) \tag{9}$$

and $P_{intravalley, \ \Gamma Y} = \left( \frac{I_{\Gamma Y ||}(\theta=90^0) - I_{\Gamma Y \perp}(\theta=90^0)}{I_{\Gamma Y ||}(\theta=90^0) + I_{\Gamma Y \perp}(\theta=90^0)} \right) \tag{10}$

The intravalley polarization degrees in ΓX ($P_{intravalley, \ \Gamma X}$) and ΓY ($P_{intravalley, \ \Gamma Y}$) directions were realized to be 95% and 96%, respectively. Such intravalley polarization values are amongst the highest polarization degrees reported, from experiments conducted at cryogenic temperatures.[236-238]

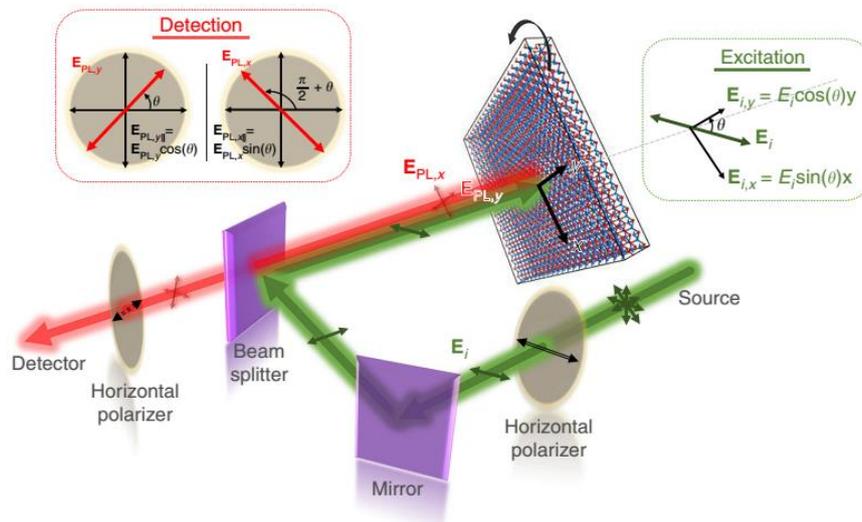

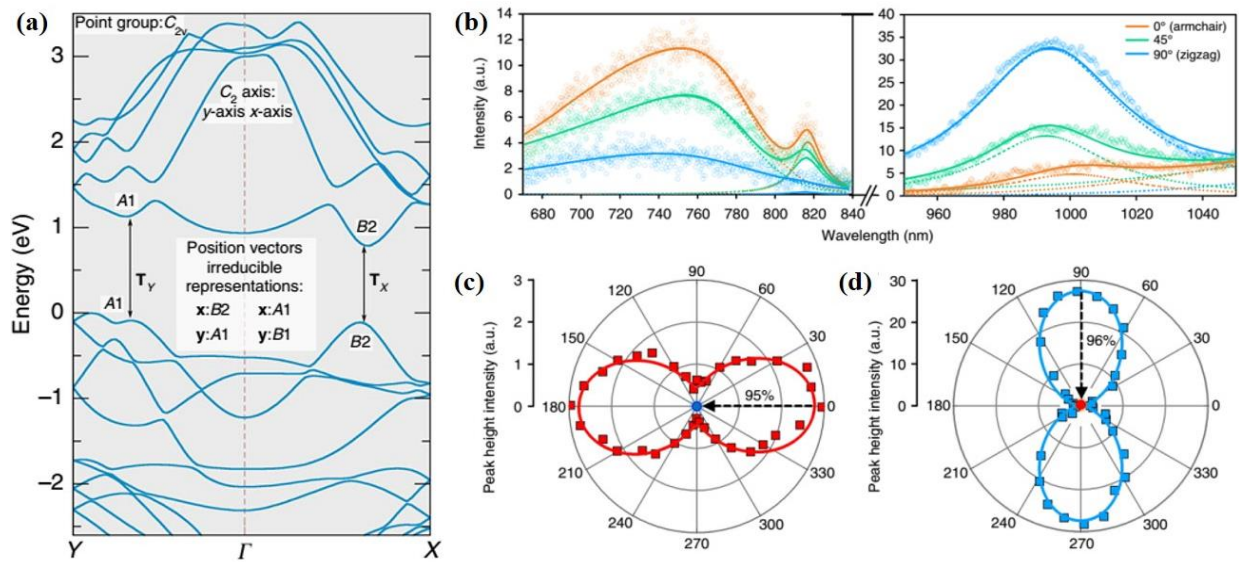

**Figure 18.** *Top panel: schematic of the experimental setup for valley polarization measurements. a) Electronic band structure of bulk SnS. The Y and X axes correspond to the armchair and zig-zag directions, respectively; PL measurements of SnS. b) Deconvolution of PL peaks with respect to sample orientation (rotation by 45º and 90º respectively) c, d) Polar plots of PL peak intensities for the different valleys (at 817 nm 995 nm). The black dashed arrows are the two data points before and after the 90° rotation of the second polarizer, corresponding to PL intensities at the ΓY and ΓX valleys respectively. Reproduced with permission[51] Copyright 2018, Author(s), licensed under a Creative Commons Attribution 4.0 License, Nature Publishing Group.*

In another study Chen et al.[50] have explored a novel valley-selective linear dichroism of the PL in SnS. The PL spectra (at 77K) of a bulk SnS (of 109 nm thickness) exhibited two photon emission peaks located at 1.16 and 1.204 eV (**Figure 19a, b**). The origin of PL was theoretically predicted previously,[37] attributed to the presence of two valleys with close energy gaps along the ΓX and ΓY directions in momentum space.[239] On the other hand, Gomez et al.[31] have reported that the $E_C^Y - E_V^X < E_C^Y - E_V^Y < E_C^X - E_V^X$, where, $E_I^J$ (J=X, Y and I=C, V) is the energy of the CBM and the VBM of the $\Gamma X$ and $\Gamma Y$ valleys, respectively. Considering that the two PL peaks arise from the band edge transitions, the polarized PL spectra were monitored and presented in **Figure 19c**. It is observed that the angular resolved PL intensities of the two PL peaks exhibited different polar shapes with $\cos^2\theta$ and $\sin^2\theta$ dependencies (**Figure 19d, e**). Furthermore, the band-edge emission along the $\Gamma X$ and $\Gamma Y$ valleys is dependent on the excitation linear polarization. In particular, for each valley $\Gamma X$ or $\Gamma Y$ the emission becomes dominant when the excitation linear polarization is parallel to its corresponding crystal orientation in real space. The experimentally obtained degrees of polarization for anisotropic emissions from $\Gamma X$ and $\Gamma Y$ valleys were 8.5% and 3.4%, respectively.

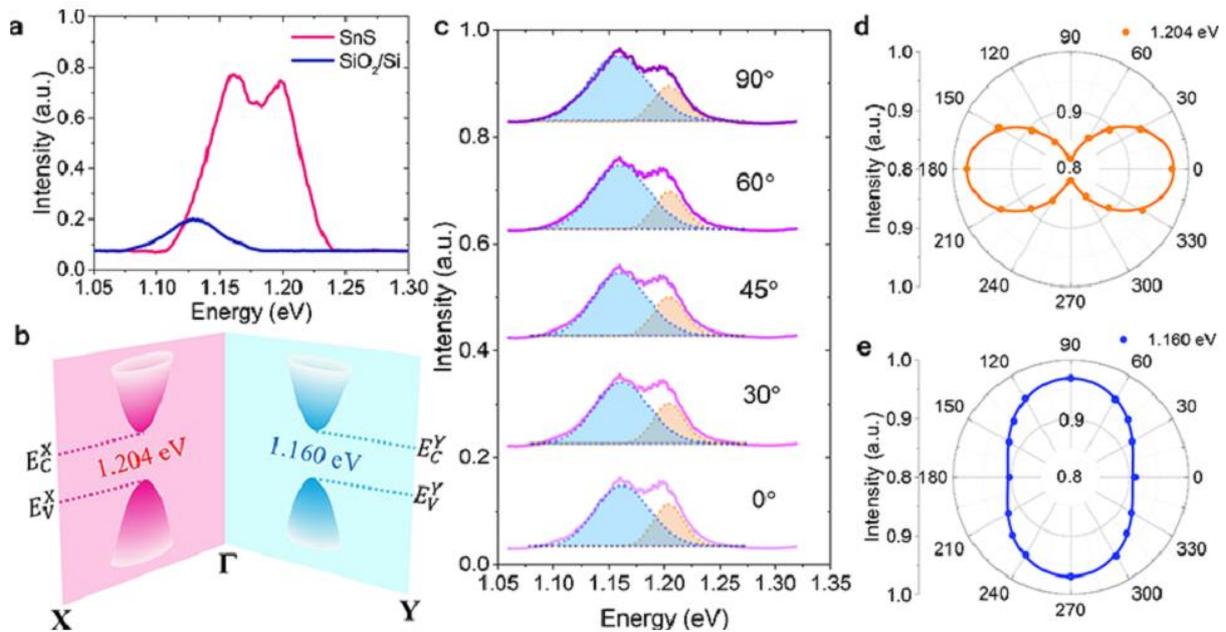

**Figure 19**. *a) PL spectra of a SnS flake and its substrate (Excitation photon energy of 2.33 eV at 77 K), b) Band structure schematics of layered SnS corresponding to two valleys of close energy gaps in the $\Gamma X$ and $\Gamma Y$ direction. c) PL spectra under different excitation linear polarization directions. d, e) Angular-resolved PL intensity of the emission peaks at 1.204 eV ($\Gamma$-X valley) and 1.160 eV ($\Gamma$-Y valley). Reproduced with permission[50] Copyright 2018, The American Chemical Society.*

## 4.4 Second Harmonic Generation

Second harmonic generation (SHG) is a nonlinear optical (NLO) process describing the light-matter interaction where, the induced polarization depends nonlinearly on the external electric field. SHG has been broadly applied for the materials analysis, sensors development, advanced spectroscopy, high-resolution imaging, lasers, frequency conversion, electro-optic modulators and switches.[240-243] Macroscopically, the nonlinear process is described by the light-induced polarization expressed as

$$P = \varepsilon_0\left(\chi E + \chi^{(2)}E^2 + \chi^{(3)}E^3 + \cdots\right) \tag{10}$$

$$P = \varepsilon_0\chi E + P_{NL} \tag{11}$$

where, $\varepsilon_0$ is the permittivity in vacuum, $\chi$ is the electric susceptibility of the medium, and E is the electric field of the incident light. The first term is the linear part, while the rest is the nonlinear part ($P_{NL}$). The $\chi^{(n)}$ n=1,2,3… correspond to the elevated-order nonlinearity in the medium. The generation of NLO response depends on the intrinsic crystalline symmetry, the microscopic transition dipole matrix, as well as on the specific frequency and orientation of the optical field applied. On top of them, a centrosymmetric or non-centrosymmetric crystal structure plays a pivotal role due to the different order electric susceptibility tensors describing each case. To date, a large number of 2D materials including mono/multilayer $MoS_2$, $MoSe_2$, $WS_2$, $WSe_2$, hBN, GaSe, and InSe has been reported to exhibit the second and higher order harmonics generation.[244-251] However, all of them belong to the centrosymmetric point group $D_{3h}$ comprising a single independent SHG susceptibility tensor element. On the other hand, the MMC monolayers belong to a non-centrosymmetric point group and thus exhibit up to five independent SHG susceptibility tensor elements. Qian and co-workers[52] have theoretically investigated the optical second harmonic generation in monolayer MMCs. In particular, the quasiparticle band structure of a monolayer GeSe has been calculated using first principles, quasi-atomic orbital method. Moreover, the second order susceptibility tensors in MMCs, MoS$_2$ and hBN, were estimated (**Figure 20b, c**). Specifically, the MMCs monolayers belong to the point group C$_{2v}$, while MoS$_2$ and hBN ones in D$_{3h}$. The D$_{3h}$ point group exhibits one independent nontrivial SHG susceptibility tensor element that satisfies the relation $\chi_{yxx}^{(2)} = \chi_{xyx}^{(2)} = \chi_{xxy}^{(2)} = -\chi_{yyy}^{(2)}$. On the other hand the C$_{2v}$ exhibits five independent susceptibility tensor elements (**Figure 20a**) satisfying the relations $\chi_{yxx}^{(2)}, \chi_{yyy}^{(2)}, \chi_{yzz}^{(2)}, \chi_{xyx}^{(2)} = -\chi_{xxy}^{(2)}$ and $\chi_{zzy}^{(2)} = \chi_{zyz}^{(2)}$. The obtained magnitude of the calculated susceptibility tensor element at 3.22 eV was calculated to be $5.16 \times 10^6$ pm$^2$/V for GeSe, which is much higher than that of MoS$_2$ ($3.02 \times 10^5$ pm$^2$/V) and hBN ($6.38 \times 10^4$ pm$^2$). Moreover, an independent SHG tensor element ($\chi_{xyx}^{(2)} = \chi_{xxy}^{(2)}$) with a substantial magnitude, which is much higher than the other three components of a GeSe monolayer, was predicted. The angular dependence of the SHG response, that can be measured for example by rotating the crystal and measure the SHG polarization and intensity can provide important information on 2D materials' anisotropy.[244, 252, 253] Considering normal incidence geometry, the angular dependent SHG susceptibilities for the point group C$_{2v}$ and D$_{3h}$ had been calculated. The corresponding angular dependent polar plots of SHG are shown in **Figure 20d-f**, together with the corresponding frequency of the excitation field, ω and the maximum χ$^{(2)}$ values. A significantly polarized colossal SHG response in monolayer GeSe with its maximum value, located an angle on 0$^o$, was observed. On the other hand, the polar plots of MoS$_2$ monolayer precisely reflect the D$_{3h}$ symmetry. In the case of monolayer GeSe, the total susceptibility ($\chi_{total}^{(abc)}$ (-2ω, ω, ω)) contains an interband contribution ($\chi_{intra}^{(abc)}$), a modification due to the intraband motion ($\chi_{inter}^{(abc)}$) and a modulation due to interband motion ($\chi_{mod}^{(abc)}$). Such contributions were accounted by six tensor terms,

$$\chi_{intra}^{(abc)} \equiv \chi_i(\omega) + \chi_i^{(v)}(2\omega) + \chi_i^{(r)}(2\omega) \qquad (12)$$

$$\chi_{inter}^{(abc)} \equiv \chi_e(\omega) + \chi_e(2\omega) \tag{13}$$

$$\chi_{mod}^{(abc)} = \chi_m(\omega) \tag{14}$$

Among such susceptibility terms, the $\chi_i^{(v)}(2\omega)$ and $\chi_i^{(r)}(2\omega)$ are the leading ones behind the giant SHG susceptibility in GeSe monolayer, governed by the interband and intraband Berry connection in the monolayer. These leading susceptibility terms are distributed at four spots in the vicinity of the first Brillouin zone and concentrated around the two valleys in the band gap. Moreover, the authors have predicted SHG strength in the monolayer and tri-layer GeSe and SnSe, which has very similar SHG strength and shape.

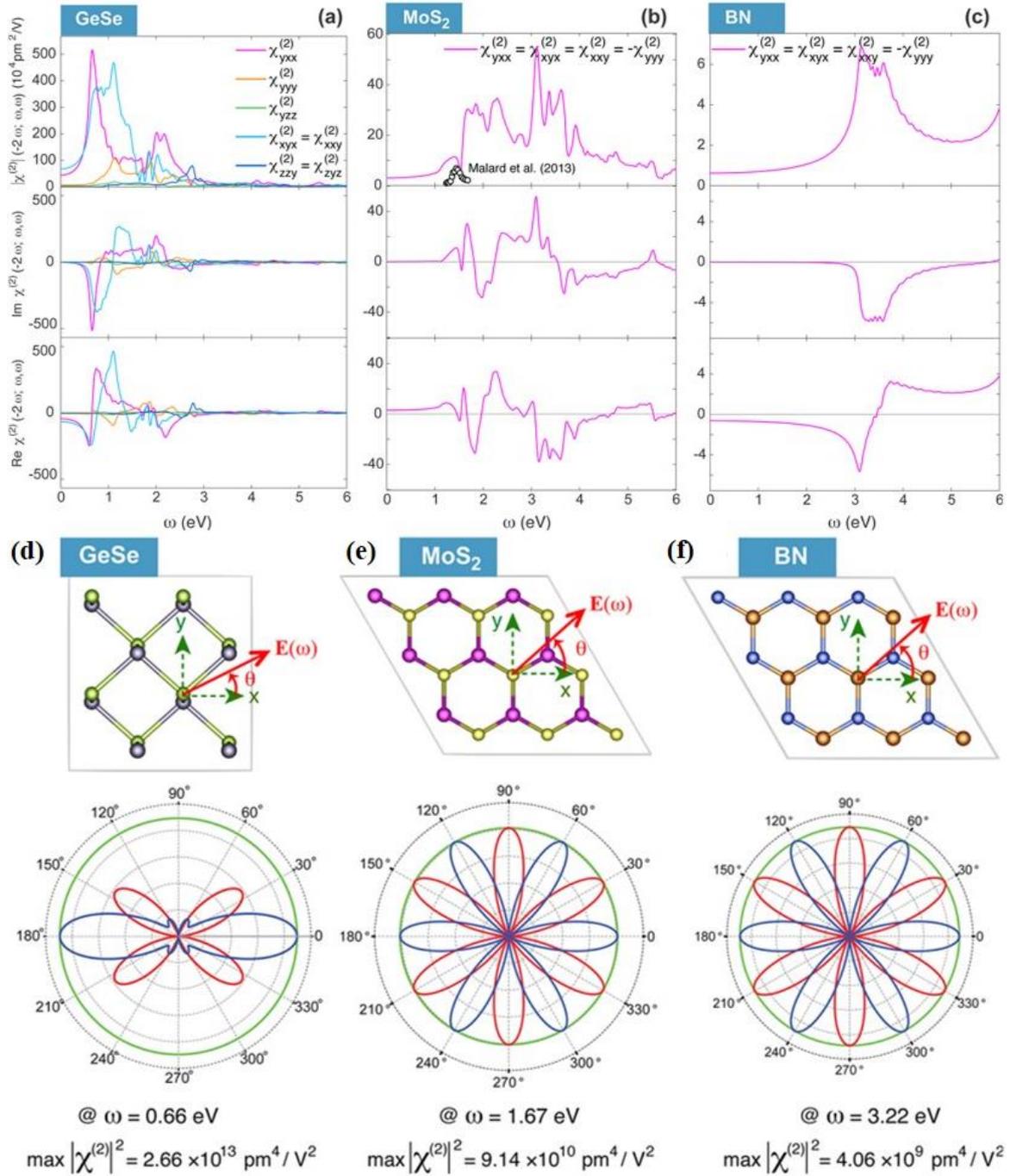

**Figure 20.** *The magnitude, imaginary and real component of SHG susceptibility for: (a) a GeSe monolayer which has seven nonzero susceptibility tensor elements. b) a MoS₂ monolayer which has four independent SHG elements. c) a h-BN monolayer with only one independent element. Black dots indicate the experimental values.; Polarization anisotropy of SHG susceptibilities in monolayer d) GeSe, e) MoS₂, and e) h-BN. The red/blue solid lines are the polarization components of the SHG response parallel/perpendicular to the polarization of the incident*

*electric field E(ω). θ is the rotation angle between E(ω) and the crystal lattice. Reproduced*

*with permission.[52] Copyright 2018, American Chemical Society.*

## 5.  Summary and Outlook

In this review article, we have presented the fundamental properties and recent advancements on the synthesis of layered 2D, in-plane anisotropic, phosphorene-analog group $IV_A$-VI MMCs and their emerging promising applications in electronics, optoelectronics and nanophotonics. The in-plane anisotropy along the armchair and zig-zag directions make these materials more important to explore the rich physics of 2D materials. Nevertheless, the experimental research on MMCs is still in a preliminary stage. Among the synthesis approaches, the mechanical exfoliation is fairly simple and easy to obtain low defect and highly crystalline nanoflakes. However, the low exfoliation yield, lateral size repeatability and reproducibility in layer number, harshly restrict its for application for large area electronics and optoelectronics. LPE is an effective method for bulk production and is widely used for the solution processable flexible organic and hybrid electronic applications. However, the control synthesis of MMCs with thickness uniformity in various solvents has not investigated yet. Besides this, the choice of appropriate solvent for successful isolation into thinner layers is also an important topic of research. In contrast, the physical and/or chemical vapor deposition is no doubt the most a promising route for large-area 2D materials production for future electronic applications. At present, the synthesized MMCs are restricted only in few monolayers and therefore the isolation of single layer with good crystallinity remains a great challenge. In general, there is a huge room for further study on the controlled synthesis of electronic-grade ultrathin, even atomically thin, MMC layer by means of top-down and bottom-up approaches.

To date, the optical and electrical anisotropic properties of orthorhombic MMCs have investigated in few-layer NSs, while the intriguing anisotropic optoelectronic response of a monolayer is only theoretically predicted. Such unique anisotropic response appeared due to the puckered crystal lattice structures of MMCs, resulting from the difference in the effective mass and refractive index along the armchair and zigzag directions respectively. On top of that, the large absorption coefficient and the large spin orbital coupling pave the way for the application of MMCs in solar energy harvesting and valley electronics. Apart from the current inability to isolate MMCs monolayers, the fast growing field of MMCs is also hampered by a huge gap between the materials' fundamental studies and device applications.

Notably, puckered structured Ge and Sn based chalcogenides are p-type semiconductors, which are desirable to form heterojunctions with n-type TMDs to explore the interface physics at the nanoscale and the subsequent electronic properties. As a consequence, strong light-matter interaction and reduced dimensionality leads to the formation of quasi-1D excitons and trions within in-plane anisotropic MMCs, while quasi-2D excitons and trions in isotropic TMDs materials. Therefore, heterostructures of in-plane anisotropic/isotropic 2D materials can provide unique interlayer interactions between quasi- 1D and 2D excitonic species. This may find promising applications including high-performance photoemitters and exciton-polariton lasers. Nevertheless, to date there are limited studies on such interlayer optical and electronic responses.[254-258] Furthermore, the highly anisotropic electrical and optical properties in MMCs integrated with 0D (quantum dots) or 1D organic and/or inorganic and perovskite semiconductor materials could open up a new path for next generation electronic applications. Overall, the very recent exciting achievements in the field of few-layer MMCs showed a great potential for their application in next generation electronic, optoelectronic and emerging nanophotonics, including valley electronics, solar cells, sensors and nonlinear optical applications. However, the wealth of arising possibilities in fundamental research of 2D MMCs and the emerging new approaches to MMCs monolayer synthesis and functionalization prescribe a future where tuning of MMC band structure and subsequent electronic properties can be accomplished with a level of sophistication that we cannot presently imagine.

*Note added.* We noticed that during the peer review process, a novel study on the PVD growth of monolayer SnS and the demonstration of room temperature purely in-plane ferroelectricity and SHG, was published by Nagashio and co-workers[259]

## Acknowledgements

This work was supported by the project MouldTex, funded by European Council framework programme HORIZON 2020, GA No-768705. The authors are thankful for the kind permission from the corresponding publishers to reproduce the figure in this article.

## Conflict of Interest

The authors declare no conflict of interest.

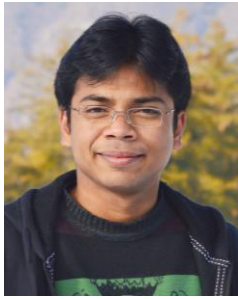

**Abdus Salam Sarkar** received his M.Sc. degree in Physics from Department of Physics and Astrophysics, University of Delhi, India and his Ph.D. degree in *emerging 2D materials optoelctronics* from School of Basic Sciences, Indian Institute of Technology Mandi, India, with *Prof. Suman K. Pal*. Currently, he is a postdoctoral research fellow in Institute of Electronic Structures and Laser, Foundation for Research and Technology Hellas, Greece with *Prof. Emmanuel Stratakis*. His research interests focused on introducing of novel 2D materials, understanding of 2D quantum science and demonstration in optics, optoelectronics, photonics, energy and biophysical electronic devices.

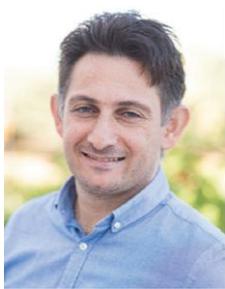

**Emmanuel Stratakis** is a research director at IESL/FORTH. He received his Ph.D. in physics and he was appointed as a visiting researcher at the University of California, Berkeley. He is currently leading the "Ultrafast Laser Micro- and Nano- processing" laboratory. His research interests are in the fields of ultrafast laser interactions with materials, nanomaterials, and related applications. He is the director of the Nanoscience Facility of FORTH, part of the NFFA-Europe EU Infrastructure. He is a national representative to the Horizon 2020 High-Level Group of EU on Nanotechnologies and a member of the Scientific Committee of COST.